\documentclass{article}

\usepackage{microtype}
\usepackage{graphicx}
\usepackage{subcaption}
\usepackage{booktabs} 
\usepackage[utf8]{inputenc}
\usepackage[T1]{fontenc}    %
\usepackage{wrapfig}
\usepackage{makecell}
\usepackage{lipsum}  
\usepackage{arxiv}
\setcitestyle{authoryear,open={(},close={)}}
\usepackage[subtle]{savetrees}
\usepackage{titling}

\usepackage{array}
\newcommand{\PreserveBackslash}[1]{\let\temp=\\#1\let\\=\temp}
\newcolumntype{C}[1]{>{\PreserveBackslash\centering}p{#1}}
\newcolumntype{R}[1]{>{\PreserveBackslash\raggedleft}p{#1}}
\newcolumntype{L}[1]{>{\PreserveBackslash\raggedright}p{#1}}

\definecolor{lue}{RGB}{135, 206, 250}
\definecolor{green}{RGB}{185, 226, 207}

\newcommand{\systemname}{\emph{SparkMe}\xspace}
\newcommand{\llmbaseline}{\emph{LLM-Roleplay}\xspace}
\newcommand{\storysage}{\emph{StorySage}\xspace}
\newcommand{\mimitalk}{\emph{MimiTalk}\xspace}
\newcommand{\interviewgpt}{\emph{Interview-GPT}\xspace}
\newcommand{\ep}{EP\xspace}
\newcommand{\ia}{IA\xspace}
\newcommand{\am}{AM\xspace}

\definecolor{lightblue}{rgb}{.8,.8,1}
\definecolor{lightred}{rgb}{1, 0.7, 0.7}

\def\shownotes{1}  
\ifnum\shownotes=1
\newcommand{\authnote}[2]{[#1: #2]}
\else
\newcommand{\authnote}[2]{}
\fi
\newcommand{\subtitle}[1]{%
  \posttitle{%
    \par\end{center}
    \begin{center}\large#1\end{center}
    \vskip0.5em}%
}

\def\arxiv{1}
\ifnum\arxiv=1

\else

\fi

\begin{document}


\title{\systemname: Adaptive Semi-Structured Interviewing for Qualitative Insight Discovery}
\author{David Anugraha, Vishakh Padmakumar, Diyi Yang\\
Stanford University\\
\texttt{\{davidanu, vishakhp, diyiy\}@stanford.edu}}
\date{October 10, 2026}
\maketitle

\begin{abstract}
Qualitative insights from user experiences are critical for informing product and policy decisions, but scaling qualitative data collection is constrained by the availability of trained interviewers. Recent work has explored using large language models (LLMs) to automate interviewing, yet existing systems lack a principled mechanism for balancing systematic coverage of predefined topics with adaptive exploration, or the ability to pursue follow-ups, deep dives, and emergent themes that arise organically during conversation. In this work, we formulate adaptive semi-structured interviewing as an optimization problem over the interviewer's behavior. We define interview utility as a trade-off between coverage of a predefined interview topic guide, discovery of relevant emergent themes, and interview cost measured by length. Based on this formulation, we introduce \systemname, a multi-agent LLM interviewer that performs deliberative planning via simulated conversation rollouts to select questions with high expected utility. We evaluate \systemname through controlled experiments with LLM-based interviewees, showing that it achieves higher interview utility, improving topic guide coverage ($+4.7\%$ over the best baseline) and eliciting richer emergent insights while using fewer conversational turns than prior LLM interviewing approaches. We further validate \systemname in a user study with $70$ participants across $7$ professions, where occupational domain experts rate it as producing higher-quality, more adaptive interviews and validate profession-specific insights not captured by the baseline. In addition, separately recruited experts in qualitative or human-centered research also find stronger responsiveness and probing in \systemname. The code and evaluation resources for \systemname are available as open source at~\url{https://github.com/SALT-NLP/SparkMe}.
\end{abstract}

\begin{figure*}[h]
    \centering
    \includegraphics[width=\textwidth]{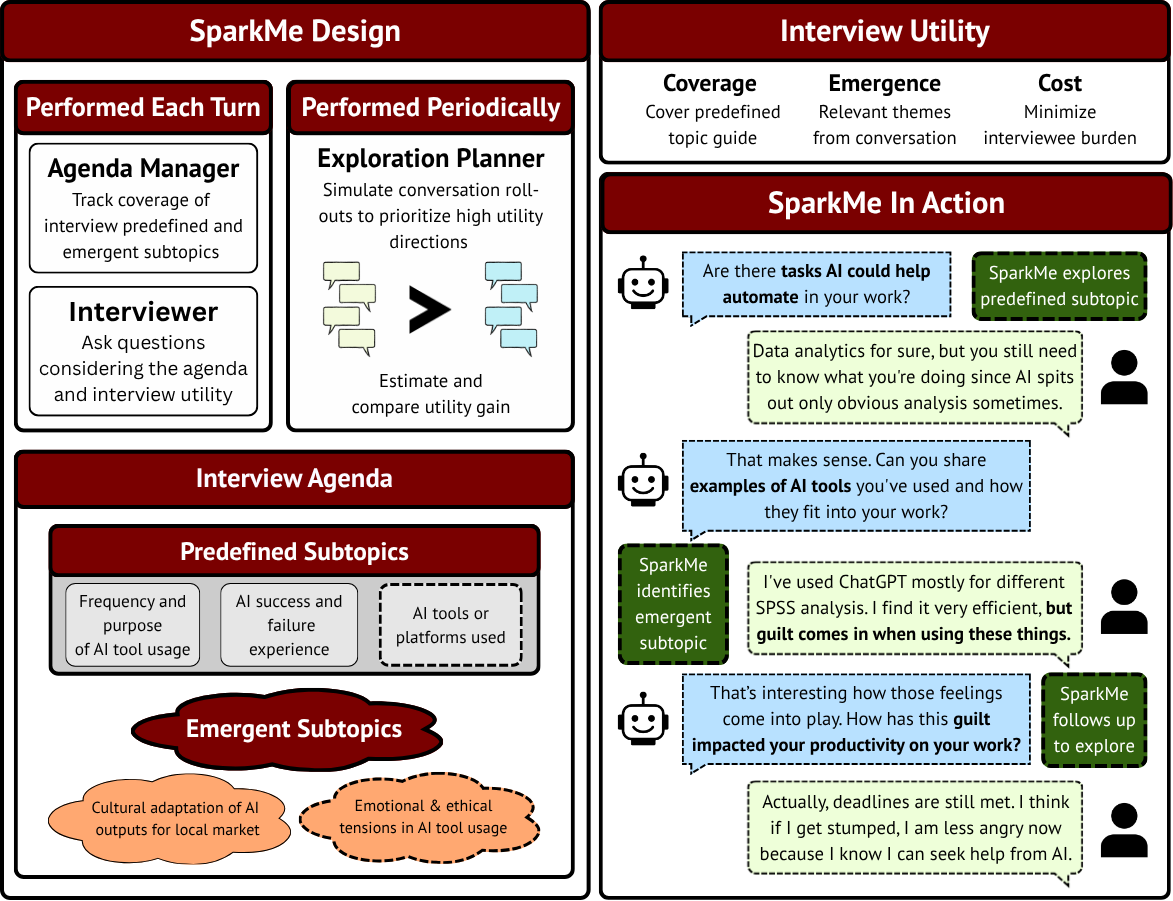}
    \caption{\textbf{Left:} \systemname optimizes interview utility by balancing coverage of a predefined topic guide with exploration of emergent themes, without unnecessarily burdening the interviewee. \systemname periodically simulates conversation rollouts to identify directions with higher expected utility gain. \textbf{Right:} An example \systemname interview from our user study on AI's impact on the workforce, focusing on the topic ``AI Experience \& Tool Adoption''. \systemname first covers a predefined subtopic (\emph{AI tools used}) and then identifies an emergent theme (\emph{emotional \& ethical tensions in AI tool usage}) that arises organically during conversation.}
    \label{fig:main-diagram}
\end{figure*}

\section{Introduction}
\label{sec:intro}

Qualitative data collection is important for understanding user experiences, practices, and perspectives~\citep{guest2013collecting}. One common approach is to use structured surveys, which can be administered to large populations of users. However, these surveys restrict responses to predefined categories, 
yielding shallow responses~\citep{patton2014qualitative, braun2013successful}. The alternative is a semi-structured interview, in which the interviewer follows a predefined interview topic guide while adapting questions based on participant responses~\citep{guest2013collecting, punch2013introduction}. This format allows interviewers to pursue directions that emerge during conversation which are relevant to the predefined topics but may not have been anticipated in the interview design~\citep{adams2015conducting, weiss1995learning}. However, conducting semi-structured human interviews at scale is constrained by the availability and time of expert interviewers.  This creates a practical bottleneck in domains that require both broad coverage and in-depth qualitative understanding, such as policy decision-making~\citep{maxwell2020value} and product design~\citep{roland2019qualitative}.

These practical limitations have inspired a line of research on using large language models (LLMs) to automate semi-structured interviews. 
Broadly, prior work can be grouped into two categories: systems that rely on a single LLM to drive the conversation~\citep{xiao2020tell, park2024generative, wei2024leveraging, liu2025understand, wuttke2025ai, handa2025interviewer}, and systems that design multi-agent architectures to coordinate interview interactions~\citep{liu2025mimitalk, cuevas2025collecting, talaei2025storysage}.

However, these methods share a common limitation that the goals of semi-structured interviewing are conveyed through prompt engineering, leaving the model to implicitly decide when to prioritize coverage of the predefined topic guide versus adaptively diving deeper into unanticipated participant responses. Since contemporary LLMs are not optimized to balance these competing goals, it remains unclear how well prompting alone can elicit the desired adaptivity \citep{schroeder2025large}. This motivates the need for a principled framework that formalizes the goals of semi-structured interviewing.

In this work, we formulate semi-structured interviewing as an optimization problem that balances coverage, exploration, and interaction cost. Inspired by prior work on task-oriented dialogue \citep{walker1997paradise}, we define a customizable objective that converts common interviewing goals into a computable framework. The objective jointly rewards (i) coverage of the predefined topic guide and (ii) discovery of emergent themes relevant to the core topics, while penalizing (iii) high interview cost, or unnecessary questions that increase interview length (\Cref{sec:interview_formulation}, \Cref{fig:main-diagram}). The relative weight assigned to each of these terms can be adjusted based on the intended interview outcome, allowing researchers to compare and select the best system for their needs.

Through the lens of this objective, we examine various LLM-based systems and find that, although they tend to cover the topic guide effectively and may permit tangential exploration, they lack explicit mechanisms that formalize and incentivize discovery and systematic coverage of emergent themes. To fill this gap, we introduce \systemname, a multi-agent system for semi-structured interviewing that performs deliberative planning to jointly optimize all components of the proposed objective (\Cref{sec:system_design}, \Cref{fig:main-diagram}). By simulating candidate conversation rollouts, \systemname estimates the expected utility of alternative interview trajectories and prioritizes lines of questioning that are likely to surface relevant subtopics not specified in the interview topic guide (\Cref{sec:coverage-planner}).

We demonstrate the value of \systemname with a case study on interviewing occupational task workers to understand their perspectives on how AI is changing the workforce \citep{shao2025future, handa2025interviewer}. 
First, in a controlled setting with synthetic user agents, 
\systemname outperforms relevant baselines on both interview topic guide coverage and overall utility, while its planning mechanism enables exploration of relevant emergent themes not anticipated by the guide (\Cref{sec:auto_experiments,sec:auto_findings}).
Then, we further validate \systemname through a user study with $70$ participants across $7$ professions (\Cref{sec:user_study}). Participants and occupational peers rate \systemname as producing higher-quality, more adaptive interviews and validate profession-specific insights not captured by the baseline. In addition, separately recruited experts in qualitative or human-centered research also find stronger responsiveness and probing in \systemname.

In all, our contributions are:
\begin{itemize}
    \item We design a tractable, customizable objective for semi-structured interviewing that balances coverage of the interview topic guide, discovery of relevant emergent themes, and interview cost.
    \item We introduce \systemname, a multi-agent interviewer that incorporates a deliberative planning step with simulated conversation rollouts to adaptively navigate this trade-off, released to the open-source community.\footnote{The code and evaluation resources for \systemname are available at~\url{https://github.com/SALT-NLP/SparkMe} along with instructions on how to customize the system to a new interview domain.}
    \item We evaluate \systemname through controlled benchmarking with simulated user agents and a human study with $70$ participants across $7$ professions. Participant ratings, peer validation, and independent evaluation by qualitative-research experts show improvements in adaptiveness, responsiveness, probing, and the elicitation of profession-specific insights.
\end{itemize}

\section{Formulating the Task of Adaptive Semi-Structured Interviewing}
\label{sec:interview_formulation}

We formalize adaptive semi-structured interviewing as a utility optimization problem over interviewer behavior.
Semi-structured interviews are a qualitative research method in which the interviewer conducts a conversation guided by a predefined topic guide, consisting of broad \emph{core topics} and more specific \emph{subtopics}, while adapting questions in response to the participant’s answers~\citep{punch2013introduction, busetto2020use, adeoye2021research, adams2015conducting}. The goal is to ensure systematic coverage of the topic guide across participants, while remaining responsive to relevant content that arises during the conversation. This adaptivity---the ability to pursue follow-up questions, conduct deep dives into promising directions, and surface emergent themes organically---is central to effective semi-structured interviewing, yet prior LLM approaches lack explicit mechanisms to optimize for it. This flexibility is critical in practice, as participants often articulate experiences, examples, or concerns that are thematically related to the predefined topics but were not anticipated when the topic guide was  created~\citep{adams2015conducting, weiss1995learning, kvale2009interviews, charmaz2008grounded}. Expert interviewers treat such participant-introduced information as integral to understanding the core topic rather than as incidental details, and follow up through open-ended probing questions~\citep{robinson2023probing}. We refer to this participant-introduced but thematically relevant information as \textbf{emergent} subtopics (i.e., content related to core topics but not specified in the original topic guide), and incorporate them into the interviewing objective to encourage appropriate probing and follow-up behavior.
At each turn, the interviewer aims to ask questions by balancing coverage of the predefined topic guide and the elicitation of emergent subtopics without unnecessarily burdening the interviewee. The relative importance of these objectives can be customized based on the goals of the researchers conducting the study.

\subsection{Preliminary Definition}
\label{sec:preliminary-definitions}

As stated earlier, the topic guide consists of a set of broad areas of interest from the user, called \emph{core topics}, each of which may be further divided into more specific \emph{subtopics}. The guide is typically derived from prior literature, previous empirical studies, or preliminary pilot data collection such as document analysis or observation~\citep{busetto2020use}. Formally, we represent this predefined list as a set of core topics
\[
\mathcal{T} = \{t_1, t_2, \ldots, t_m\},
\]
where each core topic corresponds to a high-level area of inquiry. Each core topic $t_i$ is associated with a set of subtopics
\[
\mathcal{S}_i = \{s_{i,1}, s_{i,2}, \ldots, s_{i,k_i}\}, \quad \mathcal{S} = \bigcup_{i=1}^{m} \mathcal{S}_i, 
\]
capturing more fine-grained aspects that the interviewer aims to explore. For example, a core topic such as \emph{educational background} may include subtopics like undergraduate experience, research projects, or influential mentors.

We define an \emph{emergent subtopic} as participant-introduced information that is relevant to the core topics of the interview but not captured by the predefined set $\mathcal{S}$. Let $\mathcal{S}_{emg}(R)$ denote the theoretical set of valid emergent subtopics supported by the participant responses $R$. Unlike the predefined subtopics, this set cannot be enumerated in advance and is not available to the interviewer as ground truth.

\subsection{Objective Function}
\label{sec:objective-function}

Building on prior work on task-oriented dialogue systems that maximizes task success while minimizing interaction cost~\citep{walker1997paradise,ultes2017reward,li2016deep}, we define a utility objective that captures three goals of semi-structured interviewing: systematic coverage of the predefined topic guide~\citep{guest2013collecting,patton2014qualitative}, discovery of emergent themes~\citep{charmaz2008grounded,kvale2009interviews,madill2011interaction}, and minimization of interviewee burden~\citep{weiss1995learning,adams2015conducting}.

Formally, given a sequence of questions $Q = \{q_1, q_2, \dots, q_n\}$ and the corresponding participant responses $R = \{r_1, r_2, \dots, r_n\}$, each of the metrics can be defined as follows:

\paragraph{Predefined Subtopic Coverage} The interview should have high coverage of the subtopics specified in the topic guide. $C(R \mid Q, \mathcal{S})$ measures the number of predefined subtopics $\mathcal{S}$ that are explored in sufficient depth:
\[
C(R \mid Q, \mathcal{S}) =\sum_{s \in \mathcal{S}} f_{cov}(s, R \mid Q),
\]
where $f_{cov}(s, R \mid Q) \in [0,1]$ is a real-valued score that measures the extent to which $s$ is covered in $R$. Coverage is non-deterministic and depends on responses. Importantly, a subtopic being mentioned in a subset of responses does not imply substantive coverage, as these may be brief or superficial.
We define $f_{cov}$ in a general form so that it can be instantiated to reflect the desired notion of coverage for the researcher conducting the interview.

\paragraph{Emergent Subtopic Coverage} Beyond predefined subtopics, we also reward the interviewer for discovering and covering emergent subtopics:
\[
E(R \mid Q) = \sum_{s \in \mathcal{S}_{emg}(R)} f_{cov}(s, R \mid Q),
\]
Here again, $f_{cov}$ refers to the functional form used to evaluate whether a subtopic has been covered in sufficient detail.
The key difference is that emergent subtopics are participant-driven and cannot be anticipated in advance---they are revealed organically through the participant's responses during the conversation.

\paragraph{Cost} $L(Q)$ quantifies the burden on the interviewee from responding to questions, which we seek to minimize to preserve user experience.
\[
L(Q) = \sum_{q \: \in \: Q} f_{cost}(q)
\]
where $f_{cost}$ is a function mapping each question to its cost. This can range from a fixed cost for each question to more complex functions based on the tokens needed per question or cognitive load. 

\paragraph{Overall Utility Function} Taken together, we define the utility of an interview as:
\[
U(Q, R, \mathcal{S}) = \alpha \: C(R \mid Q, \mathcal{S}) - \beta \: L(Q) + \gamma \: E(R \mid Q),
\]
where $\alpha, \beta, \gamma > 0$ are weights that control the relative importance of each term. For applications in which it is difficult to enumerate all relevant information in advance (e.g., understanding workforce disruption due to AI), the objective can place greater weight on the emergence term. In contrast, when the goal is to collect information for downstream use in rigid legacy systems, the objective can prioritize coverage of predefined subtopics. The utility is intended for comparative evaluation under a fixed weighting scheme, so absolute normalization is not required.

Since responses $R$ are stochastic and depend on both the questions asked and the participant's knowledge and communication style, the interviewer seeks the question sequence that maximizes expected utility:
\[
Q^* = \arg\max_{Q} \; \mathbb{E}_{R \sim \mathbb{P}(R \mid Q)} \big[ U(Q, R, \mathcal{S}) \big].
\]
Thus, the interviewer is optimizing over a space of question sequences $Q$ that includes decisions about which subtopics to pursue (and in what order) as well as how to pose questions, since both influence participant responses.

We note that this objective captures the primary goals of adaptive semi-structured interviewing without being exhaustive. Additional considerations, such as clarity, rapport, or ethical constraints, could be incorporated as additional terms or constraints in our objective function similarly for specific applications. 

\section{\systemname Design}
\label{sec:system_design}

\begin{table*}[!ht]
    \centering
    \resizebox{\textwidth}{!}{
    \begin{tabular}{lcccccc}
    \toprule
    \textbf{System Name} &
    \makecell{\textbf{Interview}\\\textbf{Structure}} &
    \makecell{\textbf{Stopping}\\\textbf{Condition}} &
    \makecell{\textbf{Coverage-}\\\textbf{aware}} &
    \makecell{\textbf{Deliberative}\\\textbf{Exploration}} &
    \makecell{\textbf{Open}\\\textbf{Source}} \\
    \midrule
    AI Recruiter~\citep{pathak2025ai} & Semi-structured & User-driven & \checkmark & -  & - \\
    Persona ChatBot~\citep{wei2024leveraging} & Unstructured & User-driven & - & - & - \\
    VirtualInterviewer~\citep{gomez2025virtual} & Semi-structured & Script-based & \checkmark & - & - \\
    Anthropic Interviewer~\citep{handa2025interviewer} & Semi-structured & Script-based & \checkmark & - & - \\
    Interview-GPT~\citep{wuttke2025ai} & Unstructured & User-driven & - & - & \checkmark \\
    MimiTalk~\citep{liu2025mimitalk} & Semi-structured & User-driven & \checkmark & - & \checkmark \\
    StorySage~\citep{talaei2025storysage} & Semi-structured & User-driven & \checkmark & - & \checkmark \\
    LLM-Roleplay~\citep{park2024generative} & Structured & Script-based & \checkmark & - & \checkmark \\
    \textbf{\systemname} & Semi-structured & Utility-driven & \checkmark & \checkmark & \checkmark \\
    \bottomrule
    \end{tabular}
    }
    \caption{Comparison of LLM-based interview systems. Systems vary in interview structure, stopping condition (user-, script-, or utility-driven), coverage awareness, and design mechanisms to support deliberative exploration. Unlike prior approaches, \systemname jointly balances coverage of the topic guide with exploration beyond it while minimizing interview cost.}
\label{tab:interviewer-comparison}
\end{table*}

Given the formulation of semi-structured interviewing in \Cref{sec:interview_formulation}, we examine how existing systems operationalize these objectives (\Cref{tab:interviewer-comparison}). While prior approaches introduce mechanisms to ensure coverage of the topic guide, they do not explicitly optimize for emergent discovery or formalize the exploration-coverage trade-off.\footnote{We describe these systems in more detail in \Cref{sec:related}.} To address this gap, we introduce \systemname, a multi-agent system that explicitly optimizes the proposed objective by jointly balancing topic guide coverage, discovery of relevant emergent themes, and interview cost. We adopt a multi-agent architecture for three reasons: (1) prior work suggests that a single LLM may struggle to balance these adaptive goals~\citep{schroeder2025large}; (2) simulating multi-turn outcomes in the background has proven effective in related applications~\citep{wu2025collabllm, zhang2024probing},
and (3) this approach aligns with dual-process theories of cognition~\citep{sloman1996empirical, kahneman2011thinking}, which advocate separating fast, reactive processes from deliberative long-horizon planning.

Figure~\ref{fig:main-diagram} illustrates our system, which consists of three agents that operate either at turn-level or for longer-term planning, where a turn consists of an interviewer question followed by an interviewee response. At each turn, the InterviewerAgent (\ia) conducts a conversation with the interviewee, and an AgendaManager (\am) tracks interview state and subtopic coverage (\Cref{sec:turn-level}). To encourage exploration beyond the topic guide, we introduce an ExplorationPlanner (\ep) that operates every $k$ turns. \ep evaluates future conversation directions by simulating conversation rollouts, estimating the expected utility of different exploration strategies, and providing suggestions on promising directions to pursue (\Cref{sec:coverage-planner}). These agents coordinate through a shared Interview Agenda that maintains subtopic coverage, accumulated notes, and priorities given the interview state.\footnote{The prompts for each agent action are provided in \Cref{sec:interviewer_prompts}.}

\subsection{Turn-Level Interaction}
\label{sec:turn-level}

To conduct interviews that have a natural conversational flow, we rely on two agents that run sequentially at each interview turn.

\subsubsection{InterviewerAgent (\ia)}
\label{sec:interviewer-agent}

\ia conducts a conversation with the interviewee, asking questions to cover predefined and emergent subtopics while maintaining natural dialogue flow. At each turn, the agent's context includes: (1) the Interview Agenda from \am, containing subtopic states and accumulated notes, and (2) prioritized subtopics and question suggestions from \ep. Based on the conversation history and the current context, the agent is prompted toward three general strategies, inspired by different types of follow-up questions~\citep{adams2015conducting}, to increase interview utility. These strategies do not constitute an explicit discrete action choice at each turn.

\begin{itemize}
    \item \textbf{Probe for depth}: Ask follow-up questions on the current subtopic to increase $f_{cov}(s, R \mid Q)$, thereby improving the coverage term $C(R \mid Q, \mathcal{S})$.
    \item \textbf{Explore emergence}: Pursue follow-up questions when responses introduce relevant content outside the predefined subtopic set $\mathcal{S}$, allowing the interviewer to explore candidate emergent subtopics.
    \item \textbf{Transition to next subtopic}: If the current subtopic is sufficiently covered, then move to a different uncovered or under-explored subtopic from $\mathcal{S}$, reducing any unnecessary overhead cost.
\end{itemize}

If there is no pending subtopic in the agenda, either predefined or emergent, \ia will end the interview, reducing the cost of asking unnecessary questions to the interviewee. 

\subsubsection{AgendaManager (\am)}
\label{sec:agenda-manager}

\am operates synchronously with the \ia at each turn, maintaining an Interview Agenda, which is a shared text that tracks subtopic coverage, accumulated notes, and summaries from interviewee responses. \am performs two key functions:

\paragraph{Note-taking Management.}
Based on the interviewee’s response, the agent extracts salient information and associates it with the relevant subtopic(s) in the Interview Agenda. These notes capture factual details, explanations, and insights provided by the interviewee. The agent maintains both subtopic-specific notes and topic-level notes, enabling \ia to refer to relevant context when formulating follow-up questions.\footnote{We note that recent work, \storysage, also performs similar note extraction~\citep{talaei2025storysage}. Our approach additionally organizes notes hierarchically according to the predefined subtopic structure $\mathcal{S}$, enabling explicit tracking of coverage.} 

\paragraph{Coverage assessment and summarization.} \am tracks coverage on two levels. At the subtopic level, \am estimates $f_{cov}(s, R \mid Q)$ to decide if the current subtopic $s \in \mathcal{S}_i$ is covered. If so, it is marked complete, and the corresponding notes are summarized to reduce context length while preserving key information. Similarly, at the topic level, once all subtopics within a core topic $t_i$ are complete, their summaries are further condensed into a single topic-level summary.

\subsection{Longer-Term Planning With ExplorationPlanner (\ep)}
\label{sec:coverage-planner}

\begin{figure*}[!t]
    \centering
    \includegraphics[width=\textwidth]{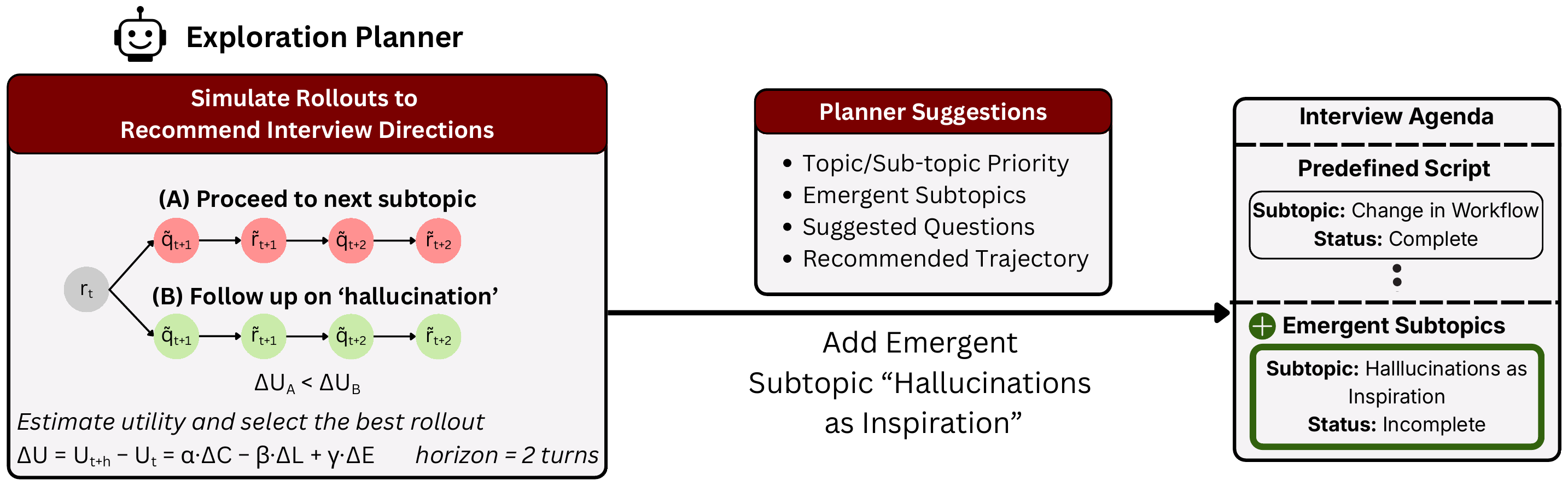}
    \caption{ExplorationPlanner (\ep) runs asynchronously every $k$ turns, simulating multiple conversation rollouts and scoring them by expected utility gain to propose conversation directions for prioritization.}
    \label{fig:coverage-planner}
\end{figure*}

Turn-level decision-making helps to efficiently cover the topic guide. However, we also wish to prioritize directions of conversation that would surface relevant emergent themes that should be explored. We design \ep explicitly for this purpose (\Cref{fig:coverage-planner}). \ep adjusts the Interview Agenda toward maximizing the overall utility function and not just the predefined subtopic coverage. Unlike \ia and \am, \ep is triggered every $k$ turns to proactively evaluate interview progress and propose question trajectories without introducing high latency.

Given the current interview state at turn $t$, with question history $Q_{1:t} = \{q_1, \ldots, q_t\}$ and response history $R_{1:t} = \{r_1, \ldots, r_t\}$, \ep aims to identify the next $h$ questions $Q_{t+1:t+h} = \{q_{t+1}, \ldots, q_{t+h}\}$ that maximize
\[
\mathbb{E}_{R_{t+1:t+h} \sim \mathbb{P}(R \mid Q_{1:t+h})}
\big[ U(Q_{1:t+h}, R_{1:t+h}, \mathcal{S}) \big],
\]
where $h$ is the planning horizon. Since direct optimization is intractable, \ep approximates this objective by simulating rollouts of conversations.

\paragraph{Conversation Rollout Prediction.}
\ep incorporates stochastic rollouts to estimate multi-turn outcomes over the horizon, which has been shown to be beneficial in related settings \citep{zhang2024probing, wu2025collabllm}. Specifically, for a candidate question sequence $Q_{t+1:t+h}$, \ep samples hypothetical interviewee responses $\tilde{R}_{t+1:t+h}$ using an LLM conditioned on the current interview context and the proposed questions.

Each rollout is scored by computing the expected utility gain:
\[
\Delta U = U_{t+h} - U_t = \alpha' \cdot \Delta C - \; 
\beta' \cdot \Delta L + \gamma' \cdot \Delta E,
\]
where $\Delta C$ is the coverage gain from newly satisfied subtopics based on simulated responses, $\Delta L$ is the turn cost incurred over the horizon $h$, and $\Delta E$ is the expected emergence gain from content outside $\mathcal{S}$. By comparing scores across multiple stochastic rollouts, \ep identifies question sequences with the highest expected utility gain. The internal planning weights $\alpha'$, $\beta'$, and $\gamma'$ are separate from the evaluation weights in \Cref{sec:objective-function} and can be adjusted according to the desired interview outcome. The simulated responses are not intended to predict what a specific participant will say, instead, they provide a lookahead heuristic for comparing possible interview directions, while \ia continues to respond to the participant's actual answers.

\paragraph{Emergent Subtopic Identification.}
Using accumulated notes in the Interview Agenda and patterns observed in $R_{1:t}$, \ep identifies candidate emergent subtopics $\mathcal{S}_{emg}'$ that are not part of the predefined set $\mathcal{S}$ but are relevant to the core topics in the interview topic guide and introduces these into the Interview Agenda.\footnote{$\mathcal{S}_{emg}'$ represents directions predicted by \ep during the interview and is distinct from the theoretical set of valid emergent subtopics $\mathcal{S}_{emg}(R)$ defined in \Cref{sec:interview_formulation}. We describe how the latter is operationalized after the interview in \Cref{sec:auto_eval}.}
\section{Experimental Validation of \systemname}
\label{sec:eval}

\systemname is designed as a general-purpose system for semi-structured interviewing that can be customized to the needs of the researcher. To evaluate its effectiveness, we conduct a case study on understanding the impact of AI on the workforce~\citep{shao2025future,handa2025interviewer}. This domain contains $|\mathcal{T}|=10$ topics and $|\mathcal{S}|=48$ subtopics and is well suited for semi-structured interviewing because practices vary across professional roles and unanticipated usage patterns frequently arise.

We first benchmark \systemname against baselines using simulated user agents as a repeatable, controlled testbed for relative comparison. Although simulated users may not fully capture the diversity of human behavior~\citep{zhou2026mind,vijayvargiya2026lost}, they allow systems and design choices to be compared under identical conditions. We then compare \systemname with the strongest baseline in a user study involving $70$ workers across $7$ professions, where participants assess their interview experiences and occupational peers validate the extracted insights. We additionally conduct a transcript-based evaluation in which experts in qualitative or human-centered research assess interviewer behavior directly.\footnote{We acknowledge the limitations of relying solely on evaluation with simulated user agents in \Cref{sec:limitations}.}

\subsection{Benchmarking \systemname With User Agents}
\label{sec:auto_experiments}

\subsubsection{Task Setup}
\label{sec:setup}

We create a suite of $200$ user agents by bootstrapping user profiles from survey responses collected from \citet{shao2025future}. Each profile contains the user’s professional background, information about their typical workflows, prior exposure, and attitudes toward AI systems.\footnote{We provide a sample user profile in Appendix \Cref{sec:app_user_profile}.}
Each user agent is instantiated with an LLM conditioned on the particular profile. During an interview, user agents are prompted to respond to questions using only the information provided in their profile, continuously maintaining a history of previous turns. 
The goal for the interviewer system is to elicit the information from the user profile to maximize topic coverage and overall interview utility, following the formulation in \Cref{sec:interview_formulation}.\footnote{The prompts for each step of \systemname are provided in \Cref{sec:interviewer_prompts}.} We provide the topic guide used for the task in \Cref{sec:app-interview-guide-list}.

For the main results, we use \texttt{Qwen3-30B-A3B-Instruct-2507} as the backbone for both the interviewer systems and user agents. We provide additional experiments using \texttt{GPT-4.1-mini} and different combinations of interviewer and user-agent models in \Cref{sec:appendix_results}. The user-agent prompt can be found in Figure~\ref{prompt:user-agent-template}.

\subsubsection{\systemname Setup}
\label{sec:our-setup}

Since ground-truth subtopic information is not available during the interview, \systemname approximates $f_{cov}(s, R \mid Q)$ using an LLM-as-a-judge prompt designed to follow the STAR heuristic (Situation, Task, Action, Result).\footnote{STAR is a structured method commonly used to elicit detailed, behaviorally grounded responses in interviews~\citep{levashina2014structured}.} By focusing on these dimensions, the \am can estimate whether the current subtopic has been adequately explored and \ep can calculate estimated utility correctly.

\systemname also defines emergent subtopics as those that (1) clearly fall within an existing interview topic, (2) do not belong to any existing subtopic under that topic, (3) enable a qualitatively new line of inquiry rather than deepening an existing subtopic, and (4) reveal at least one new dimension, pattern, or tradeoff.

Regarding system hyperparameters, unless otherwise specified, for all reported results, \systemname uses weights of $\alpha' = \beta' = \gamma' = 1$, and plans $3$ rollouts every $k = 2$ turns with a horizon of $h = 3$. Further experiments, such as ablation on different hyperparameters and complete implementation details, including prompts for coverage and emergence judgments, are included in \Cref{sec:appendix_expt_details} and \Cref{sec:interviewer_prompts}.

\subsubsection{Baselines}
\label{sec:baselines}
We compare the agent design of \systemname (\Cref{sec:system_design}) with the following baselines:

\begin{itemize}
    \item \storysage \citep{talaei2025storysage}, a multi-agent interviewer system that is designed to conduct biographical interviews. \storysage also follows a modular breakdown of question planning and interviewing after consuming a list of topics as input. \storysage also allows optional tangential questions when new themes emerge or when user enthusiasm is detected, as specified in its prompting strategy. Since there is no specific stopping rule in \storysage, we limit the number of turns to be $72$, which is $1.5\times$ the number of subtopics we have. \Cref{sec:appendix_storysage} documents the prompts and experimental setup used in our implementation.
    \item \llmbaseline, based on the interviewer in \citet{park2024generative}, is a single-LLM baseline that is prompted to perform the role of an interviewer.\footnote{We note that Anthropic Interviewer~\citep{handa2025interviewer} describes an interviewer system that relies on system-level instructions specifying qualitative interview best practices and an interview topic guide. Both \llmbaseline~\citep{park2024generative} and \interviewgpt~\citep{wuttke2025ai} instantiate open-sourced variants of this paradigm.} This interviewer proceeds through the interview topic list sequentially, asking questions aligned with each subtopic and re-asking or refining questions when the responses are insufficient or incomplete. More details about the implementation of this baseline can be found in \Cref{sec:appendix_llm_baseline}.
    \item \interviewgpt~\citep{wuttke2025ai}, which is a single-agent interviewer that explores different topics in the interview topic guide dynamically while applying interview guidelines derived from qualitative research literature, such as active listening and probing for more insights when answers are too short or unclear~\citep{adams2015conducting}. Unlike \llmbaseline, which explicitly traverses the interview topic guide sequentially, \interviewgpt relies entirely on prompt-level guidance to determine which topics to explore and when, without an explicit subtopic traversal strategy. Similarly, since there is no specific stopping rule in \interviewgpt, we limit the number of turns to $72$. More details about the implementation of this baseline can be found in  \Cref{sec:appendix_interviewgpt_baseline}.
    \item \mimitalk~\citep{liu2025mimitalk}, a multi-agent interviewer system in which one LLM provides strategic oversight and constraint enforcement while a second LLM generates interview questions and dialogue within those constraints. Similarly, since there is no specific stopping rule in \mimitalk, we limit the number of turns to $72$. More details about the implementation of this baseline can be found in \Cref{sec:appendix_mimitalk_baseline}.\footnote{\mimitalk represents a strong baseline that has been \href{https://mimitalk.app/}{commercialised}.}
\end{itemize}

\subsubsection{Evaluation Metrics}
\label{sec:auto_eval}

To compare \systemname to the baselines, we evaluate the coverage and utility of these methods on the suite of user agent interviews 
(\Cref{sec:interview_formulation}).

\paragraph{Predefined Subtopic Coverage.}
We evaluate coverage $f_{cov}$ using an LLM-as-a-judge that assigns a Likert score from 1--5, a commonly used evaluation paradigm in prior LLM-as-a-judge work~\citep{kim2023prometheus, anugraha2025r3}, to indicate whether a subtopic $s \in \mathcal{S}$ is addressed in the interview transcript. The judgment is made by comparing the transcript against ground-truth information for each subtopic derived from the user profile. We provide the prompts used to score predefined subtopics in~\Cref{sec:prompt_subtopic_coverage_evaluation}, and verify alignment between LLM judgments and expert human judgments in~\Cref{sec:coverage_llm_as_judge}, where LLM judgments achieve a Kendall’s $\tau$ of $0.52$ with human judgments.

\paragraph{Emergent Subtopic Coverage.}
Because no ground-truth set of emergent subtopics is available in advance, we operationalize $\mathcal{S}_{emg}(R)$ after each interview. We apply the same post-hoc identification procedure to the completed transcript from every system and then evaluate whether each identified subtopic is covered in sufficient detail. This procedure is applied independently of \ep and does not use the candidate set $\mathcal{S}_{emg}'$ predicted during the interview. We provide the prompts used to identify and score emergent subtopics in~\Cref{sec:prompt_subtopic_coverage_evaluation}.

\paragraph{Cost.}
Interview cost, $f_{cost}$, is measured with a piecewise cost function in which the cost is $0$ for the first $|\mathcal{T}| = 10$ turns, where $10$ is the number of core topics in the interview topic guide. Additional questions result in a cost of $1$ each.

\paragraph{Utility.}
We compute overall interview utility by combining coverage of predefined and emergent subtopics and cost, following the formulation in \Cref{sec:interview_formulation}. We set $\alpha = \frac{1}{|\mathcal{S}|} = \frac{1}{48}$, $\beta = \frac{1}{1.5|\mathcal{S}|} = \frac{1}{72}$, and $\gamma = \frac{2}{|\mathcal{S}|}= \frac{1}{24}$, reflecting the need to encourage exploration while keeping the range of utility between $-2$ and $2$.\footnote{Our formulation of interview utility can be customized by the needs of the researcher. We show the impact of ablating the relative weights for the factors in \Cref{sec:appendix_results}.}

\paragraph{Interview coherence and flow.}
We measure if the questions are asked in a meaningful sequence using an LLM-as-a-judge rubric. Grounded in work on conversational analysis, each transcript is scored on the dimensions of (a) local coherence, or the sequential connection between successive questions, (b) transition quality, or the smoothness of subtopic shifts, and (c) contingent responsiveness, or the grounding of follow-up questions in prior responses \citep{sacks1974simplest, schegloff2007sequence, ongena2006methods}. We provide more discussion about these dimensions and the prompt used in \Cref{sec:appendix_readability}.\footnote{We use LLM-as-a-judge as a soft measure of interview coherence and flow in the automatic analysis. This is further validated in the user study in \Cref{sec:user_study}.}

\paragraph{Question complexity.}
Inspired by \citet{wuttke2025ai}, we evaluate if the different methods generate questions that are qualitatively different in form from one another with readability metrics of Flesch-Kincaid reading level and ease \citep{flesch1948new, kincaid1975derivation} as well as a proxy for question complexity. We want to identify if different methods obtain coverage of subtopics more efficiently by asking more challenging questions, which might hinder user experience.

\subsubsection{Findings}
\label{sec:auto_findings}

\begin{figure*}[!t]
\centering
    \includegraphics[width=\textwidth]{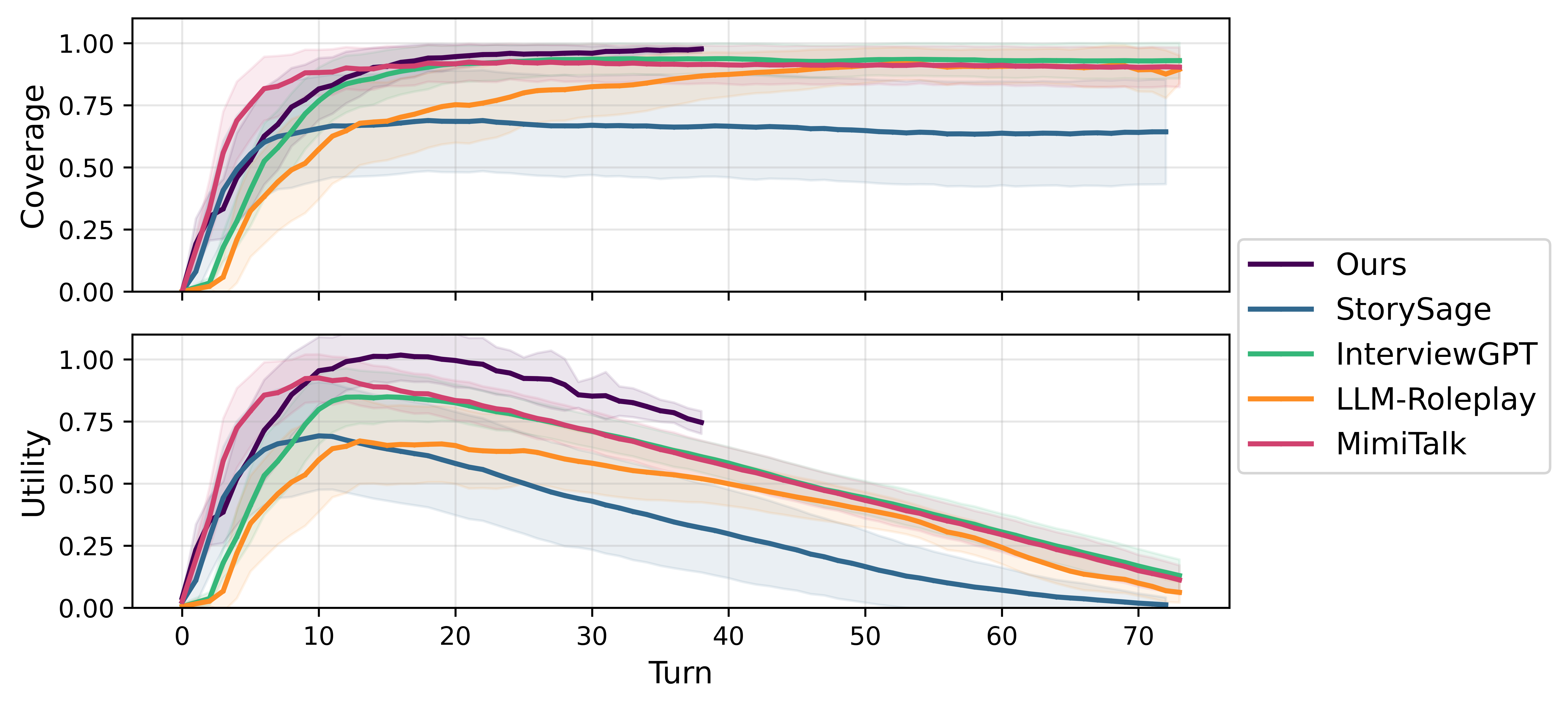}
    \caption{Predefined subtopic coverage and utility of subtopics ($y$-axis) as a function of the number of interview turns ($x$-axis) for different systems (\Cref{sec:baselines}), using \texttt{Qwen3-30B-A3B-Instruct-2507} as the backbone for the interviewer LLM and user agent. 
    \systemname (Ours) efficiently converges to a higher coverage and utility value, consistently outperforming other baselines (\Cref{sec:auto_findings}).}
    \label{fig:coverage-utility}
\end{figure*}

\begin{figure*}[t]
    \centering
    \includegraphics[width=0.95\textwidth]{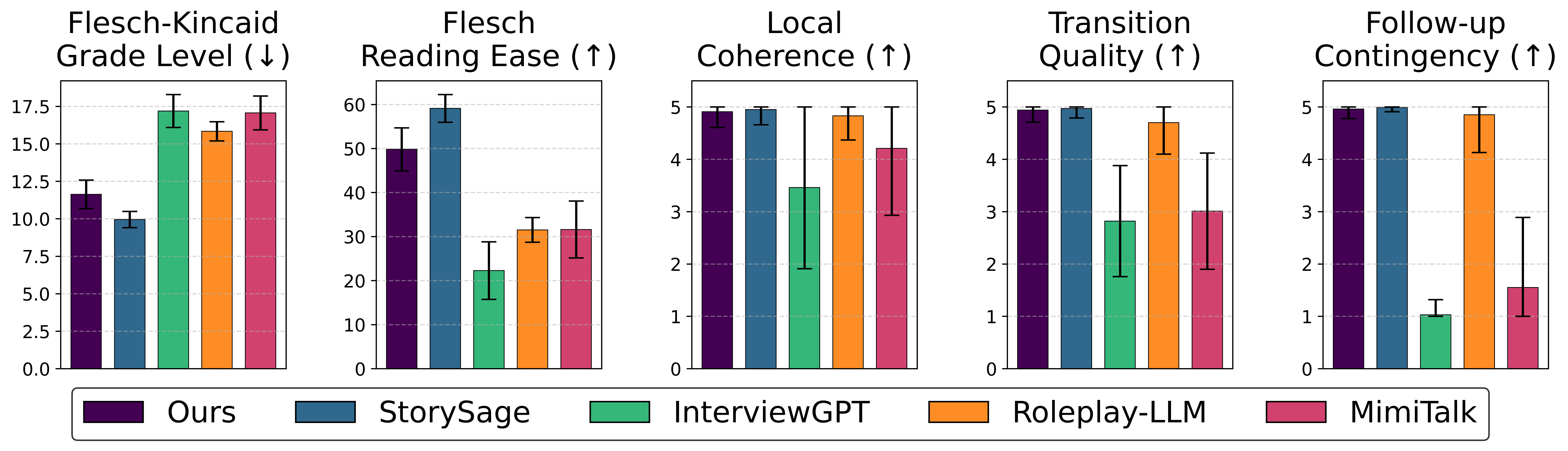}
    \caption{The first two panels show the Flesch-Kincaid Grade Level and Flesch Reading Ease of the questions in the automated evaluation. The last three panels show local coherence, transition quality, and follow-up contingency of the interview questions, rated on a 1–-5 Likert scale, to evaluate the overall quality and flow of the interviews.}
    \label{fig:readability-flow}
\end{figure*}

\begin{figure*}[t]
    \centering
    \includegraphics[width=0.8\textwidth]{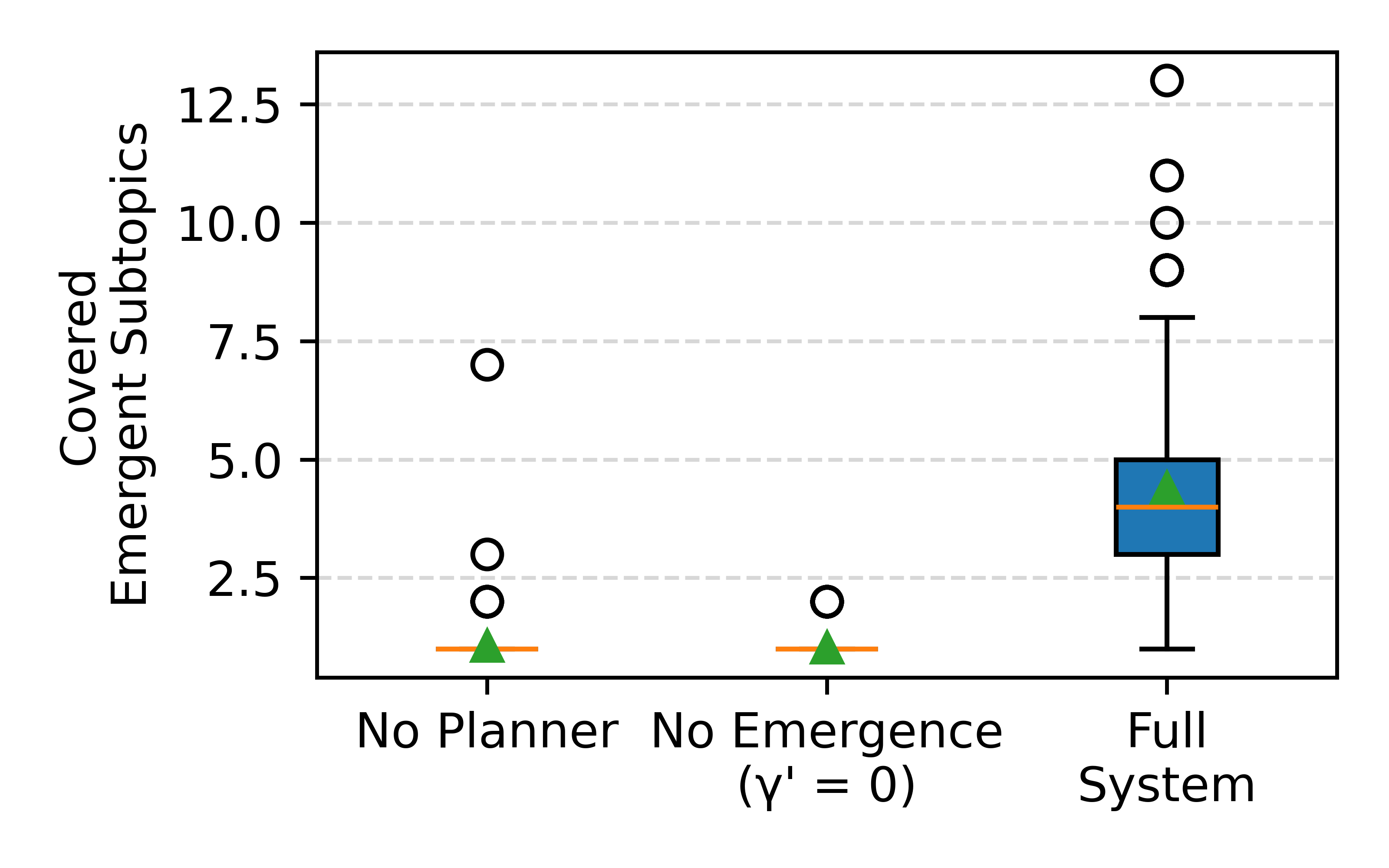}
    \caption{Number of emergent subtopics identified and covered by \systemname with and without \ep. The number of covered emergent subtopics for other baselines is not shown since they have 0 covered emergent subtopics according to our evaluation setup.}
    \label{fig:emergence-box}
\end{figure*}

\begin{figure*}[!t]
    \centering
    \includegraphics[width=\textwidth]{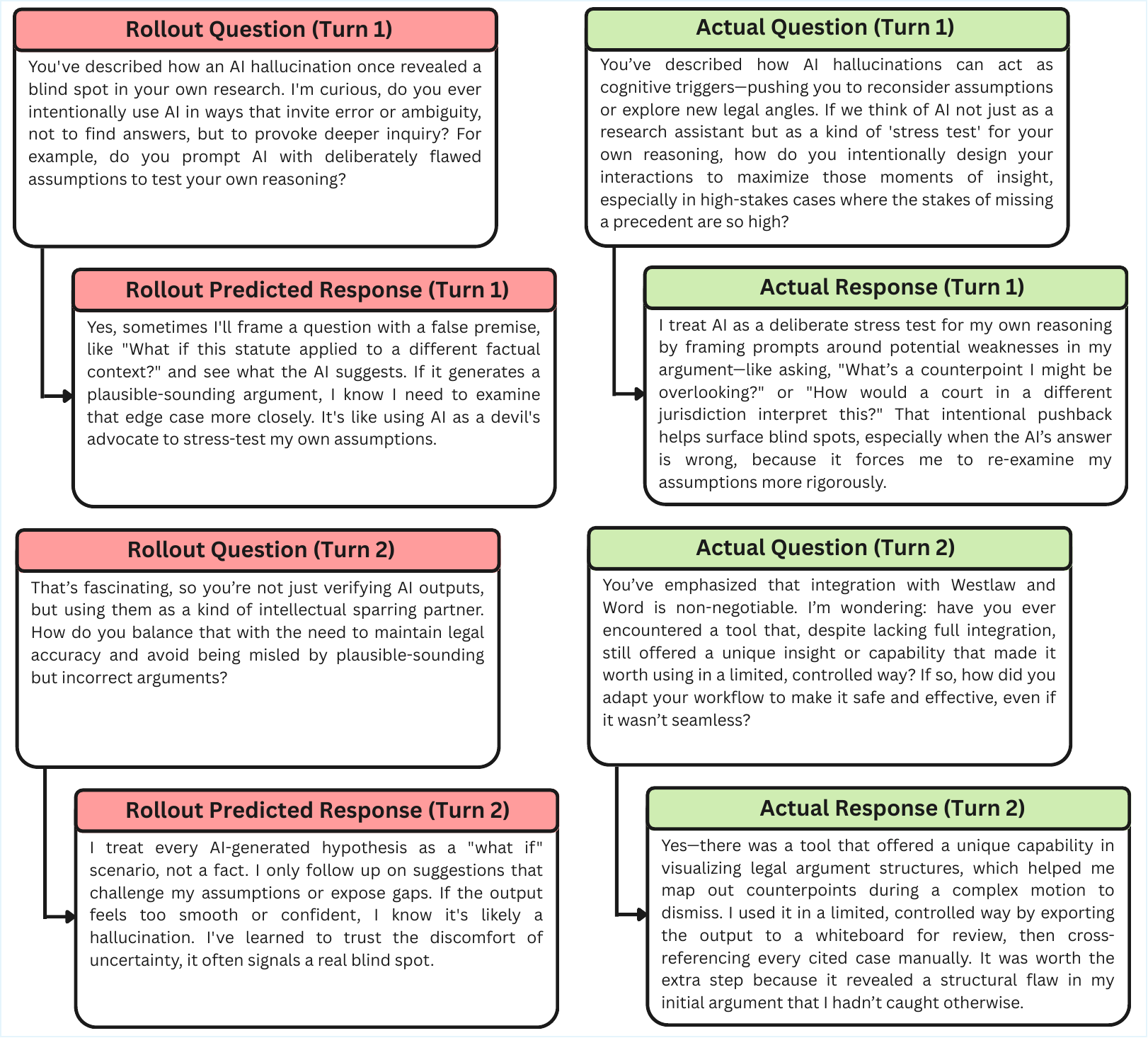}
    \caption{The left conversation illustrates a high-utility simulated rollout generated during planning to prioritize emergence, while the right conversation shows the resulting real interaction. The \ep selects the next question to explore \emph{stake-based triage of AI use}, which is identified as a high-potential emergent subtopic based on partial coverage and estimated emergence value.}
    \label{fig:sample-rollout}
\end{figure*}

\paragraph{\systemname reaches the highest final coverage and peak utility.}
Since coverage accumulates over the conversation, we report each interview's final coverage and average it across user agents. In contrast, utility includes a per-turn cost and may decline after additional questions cease to provide sufficient information gain, so we report the peak of the mean utility trajectory. From \Cref{fig:coverage-utility}, \systemname consistently outperforms all baselines in both predefined subtopic coverage and overall utility. \systemname achieves the highest average coverage ($0.977$) within $38$ turns, outperforming \storysage ($0.643$), \interviewgpt ($0.930$), \llmbaseline ($0.894$), and \mimitalk ($0.903$). A similar trend holds for utility, where \systemname attains the highest peak score ($1.017$) versus \mimitalk ($0.882$), \interviewgpt ($0.807$), \storysage ($0.656$), and \llmbaseline ($0.636$). Note that these trends hold when using \texttt{GPT-4.1-mini} as the backbone or when using different user agent models, shown in \Cref{sec:appendix_results}. Since the magnitude of the gains varies across backbones, interviewer-model selection should be treated as a deployment-specific hyperparameter and evaluated for the target setting.

\paragraph{Multi-agent-based systems obtain coverage efficiently but require careful design.}
From \Cref{fig:coverage-utility}, we see that \storysage, \systemname, and \mimitalk achieve predefined subtopic coverage and corresponding utility at a faster rate than single-agent baselines such as \llmbaseline and \interviewgpt. However, as the number of turns increases, \llmbaseline eventually outperforms \storysage, converging to a higher average coverage. This suggests that while modular, multi-agent-based systems can efficiently cover predefined topics, their final performance is sensitive to agent design, underscoring the need for careful evaluation prior to deployment.

\paragraph{Simulating conversation rollouts allows \systemname to explore relevant emergent subtopics.} From Figure \ref{fig:emergence-box}, we see that on average, incorporating \ep into \systemname leads to the discovery of more emergent subtopics across the automatic evaluation. We note that, as intended based on their design, none of the baselines cover emergent subtopics under this automatic measure. Figure~\ref{fig:sample-rollout} illustrates how a simulated rollout generated by \ep helps identify potential emergent subtopics.\footnote{\Cref{fig:coverage-utility-comparison-horizon} in \Cref{sec:appendix_results} has results when varying the number of turns in each rollout.}

In this example, \ep determines that the user’s perspective on AI hallucinations, specifically their ability to act as cognitive triggers, could lead to high-impact conceptual emergence. Based on this, the rollout plan aims to explore the user’s insight into \emph{stake-based triage of AI use}, focusing on how AI can be intentionally used in low-risk settings to probe uncertainty and stress-test ideas. The interviewer follows this plan by steering the conversation toward how AI risks are managed in high-stakes legal contexts. Although the user’s response does not match the predicted answer in terms of specific tools discussed, the interviewer maintains the same thematic direction. The interviewer then builds on earlier parts of the conversation by connecting the discussion to tool separation and safeguards used to manage AI risk, allowing for a deeper examination of the user’s approach to isolating exploratory AI outputs from authoritative decision-making. Further examples are provided in \Cref{sec:examples_auto_eval_emergence}.

Here we mainly demonstrate the ability of \systemname to explore beyond the topic guide. We confirm that the relevance and value of insights obtained from interviews with human participants are evaluated separately in \Cref{sec:user_study}.\footnote{We note the limitations of using LLM-based user agents in \Cref{sec:limitations}.}

\paragraph{\systemname obtains high coverage without asking overly complicated questions while maintaining interview coherence.} From \Cref{fig:readability-flow}, the Flesch–Kincaid Grade Level indicates that \systemname asks questions at an early college reading level, suggesting that the questions are more complex than \storysage’s simpler phrasing, while remaining more readable than \llmbaseline, \mimitalk, and \interviewgpt. This suggests that despite optimizing for coverage, \systemname maintains an accessible question complexity. \systemname also scores highly on all three coherence measures, suggesting that the gains in efficient coverage of predefined subtopics (\Cref{fig:coverage-utility}) and in following up on emergent subtopics (\Cref{fig:emergence-box}) do not come at the expense of question accessibility or interview coherence under these automatic measures. We note that the baselines, in particular \storysage and \llmbaseline, also perform well on these metrics, which is expected given the strong capabilities of the underlying LLMs.\footnote{\mimitalk and \interviewgpt tend to score lower on coherence, as they frequently repeat questions once the script is exhausted; we discuss this behavior further in \Cref{sec:appendix_results}.}

\subsection{Human Evaluation of \systemname}
\label{sec:user_study}

Having observed the strong performance of \systemname in the synthetic setting compared to the baseline systems, we now proceed to evaluate the quality of insights obtained from interviews through a user study. 
We compare \systemname to \mimitalk, the strongest baseline system in the automatic evaluation. We conduct interviews following the same interview topic guide as in \Cref{sec:auto_experiments}, provided in \Cref{sec:app-interview-guide-list}, with the aim of understanding how workers across different fields use AI tools in their daily workflows.

We recruited $70$ occupational domain experts across seven professions as participants for the interview study via Upwork\footnote{\url{https://www.upwork.com/}}. These participants were randomly assigned to either \systemname or \mimitalk, with $35$ participants in each system and $5$ participants from each of the following professions: researchers, software engineers, HR or administrative staff, creative and content professionals, educators, data analysts or scientists, and business or supply chain operations. Each participant completed a single interview session lasting up to $45$ minutes, followed by a post-interview survey, and was compensated at a rate of USD \$20--\$30 per hour. Interviews were conducted through a web-based interface, and participants could respond using either text or audio, according to their preference.\footnote{We provide further detail about recruitment in \Cref{sec:upwork}.} The study procedure was approved by an Institutional Review Board (IRB), and the interview protocol was designed to avoid the collection of personally identifiable information (PII), with explicit instructions discouraging participants from sharing identifying details during the interview.\footnote{The user study data is not publicly released to protect participant privacy. Researchers interested in accessing the data for replication or follow-up studies may contact the authors.}

\subsubsection{Evaluation Protocol}

To qualitatively evaluate the systems, we conducted a post-interview questionnaire in which participants reflected on their interviews along two main dimensions: (1) participant-perceived interview quality and (2) peer validation of emergent insights. Second, experts in qualitative or human-centered research assess interviewer behavior directly from a system-anonymized subset of full transcripts (\Cref{app:sec-hci-expert-eval-results}). All ratings were collected on 5-point Likert scales, with the detailed definitions of these rubrics provided in \Cref{sec:app-eval-rubrics,sec:app-hci-eval-rubrics}. Finally, as an additional robustness check, we also conducted an analysis to assess the presence of leading or suggestive questions and found only minimal instances across both systems (\Cref{sec:leading-question-eval}).

\paragraph{Participant Self-Assessment.} In the first part of the questionnaire, participants were provided with the breakdown of topics and subtopics discussed in the interview, with the summaries serving as a structured reference for recalling what was discussed. We normalized the summary format across both systems to minimize presentation differences. These are then rated along two dimensions:
\begin{enumerate}
    \item \textbf{Content Quality.} This includes coverage, depth, and correctness/accuracy of the information presented, reflecting how well the system captured relevant details across topics ($C(R \mid Q, \mathcal{S})$).
    \item \textbf{Interaction Experience.} This includes clarity of questions, adaptiveness to responses, and comfort during the interview.
\end{enumerate}

\paragraph{Peer Validation of Emergent Insights.} In the second part of the evaluation, participants from the same professional field reviewed emergent content extracted from their peers’ interviews to evaluate the quality of emergent content $\mathcal{S}_{emg}(R \mid Q)$. The goal is to evaluate the value of insights obtained on subtopics that were not anticipated in the script. The insights are identified using the post-hoc procedure described in \Cref{sec:auto_eval}. For each insight, participants rated the relevance to the corresponding topic, as well as its topic-level emergence, or its distinctiveness relative to other subtopics within the same topic. Next, participants rated the industry-level value of these insights, which measures the overall analytical contribution of the insights, including both surprise and relevance to the field.

\paragraph{Expert Evaluation of Interviewer Behavior.}
To assess interviewing practice directly, five experts in qualitative or human-centered research review 20 de-identified transcripts, comprising 10 interviews from each system, with system identities hidden. Each transcript receives two independent ratings. The experts evaluate preservation of participant authorship, contingent responsiveness, and probing quality using five-point scales. Complete recruitment, assignment, analysis, and agreement details are provided in \Cref{app:sec-hci-expert-eval-results}.

\subsubsection{Results}

\begin{figure}[!t]
    \centering
    \includegraphics[width=0.75\columnwidth]{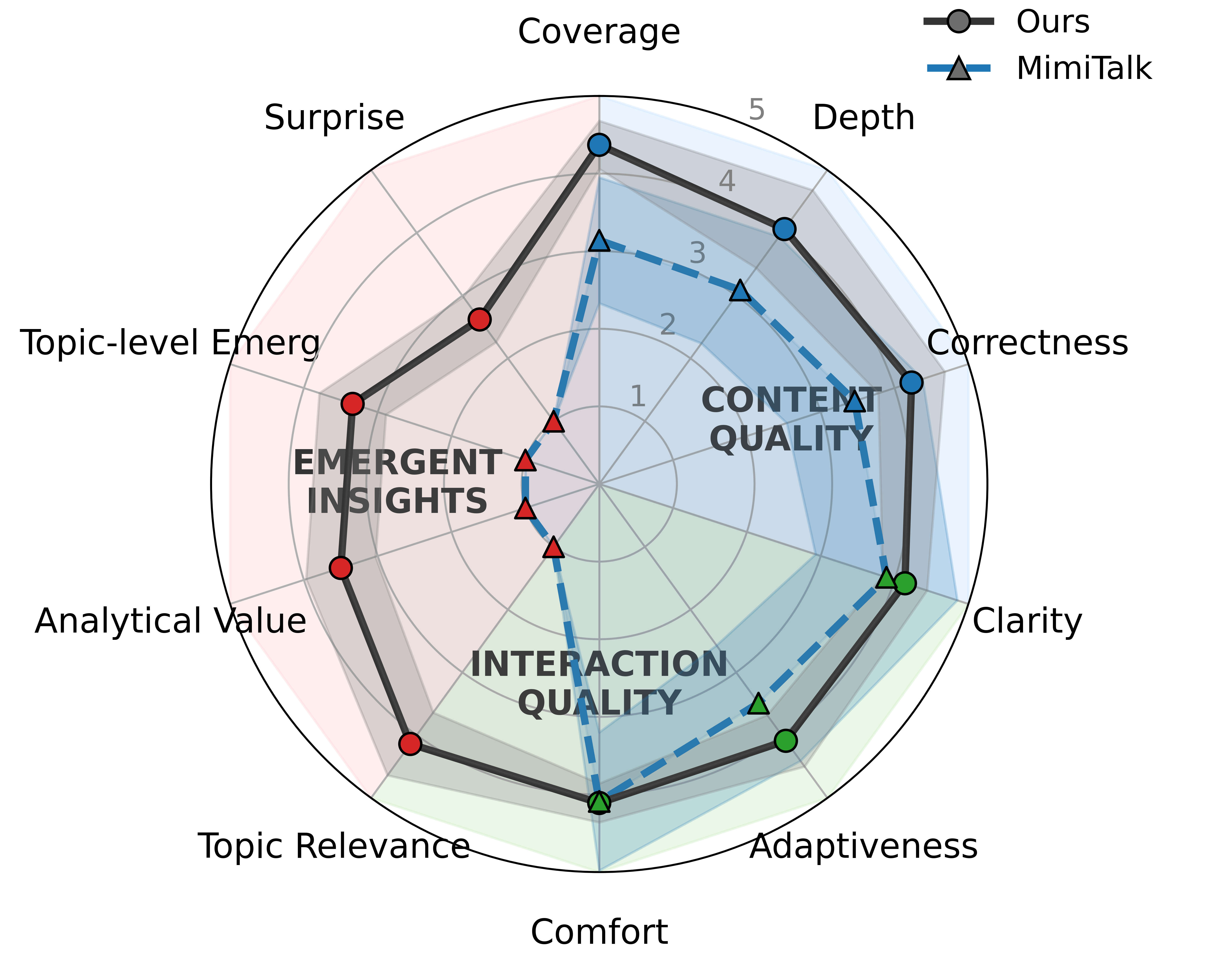}
    \caption{Overall human evaluation of \systemname and \mimitalk across three dimensions (5-point Likert scale). Content quality (blue) is evaluated based on coverage, depth, and correctness of information. Interaction quality (green) is evaluated based on clarity, adaptiveness, and conversational comfort. Emergent insights (red) is evaluated based on topic relevance, topic-level emergence, surprise, and analytical value of newly surfaced ideas.}
    \label{fig:overall-human-exp}
\end{figure}

Figure~\ref{fig:overall-human-exp} summarizes the user experience ratings across different qualitative measurements.

\paragraph{Participants score \systemname highly on content coverage, depth, and correctness.} Figure~\ref{fig:overall-human-exp} shows that \systemname is rated consistently higher than \mimitalk across all content-quality dimensions: coverage ($4.37 \pm 0.31$ vs.\ $3.14 \pm 0.81$), depth ($4.06 \pm 0.62$ vs.\ $3.09 \pm 0.85$), and correctness ($4.23 \pm 0.45$ vs.\ $3.46 \pm 0.92$), stable across professions and consistent with automated evaluation results (\Cref{sec:auto_experiments}). All three differences are statistically significant ($p < 0.001$; \Cref{tab:sparkme_mimitalk}).

\noindent \textbf{\systemname is perceived as more adaptive, while both systems remain clear and comfortable.}
Participants rate both systems similarly on clarity ($4.14 \pm 0.30$ vs.\ $3.89 \pm 0.96$, $p=0.149$) and comfort ($4.11 \pm 0.25$ vs.\ $4.09 \pm 0.89$, $p=0.899$). However, \systemname receives significantly higher adaptiveness ratings ($4.09 \pm 0.41$ vs.\ $3.49 \pm 0.92$, $p < 0.001$). This confirms that despite the additional planning and coordination complexity to be more adaptive, \systemname maintained a natural conversational flow during interviews.

\noindent \textbf{Qualitative-research experts find stronger responsiveness and probing for \systemname.} Experts in qualitative or human-centered research rate \systemname higher for contingent responsiveness ($4.00$ vs.\ $2.95$) and probing quality ($3.85$ vs.\ $2.70$), with both differences remaining significant after Holm correction ($p_{\mathrm{Holm}}=0.047$). Preservation of participant authorship does not differ significantly between systems ($3.75$ vs.\ $4.20$, $p_{\mathrm{Holm}}=0.266$). These assessments complement participants' higher ratings of \systemname's adaptiveness and depth. Complete results and agreement statistics appear in~\Cref{app:sec-hci-expert-eval-results}.

\paragraph{\systemname elicits relevant and valuable emergent insights beyond the predefined guide list.} For peer validation of emergent insights, \systemname obtains insights that are highly relevant to the core topic ($4.14 \pm 0.50$) with moderate topic-level emergence ($3.34 \pm 0.45$). Thus, \systemname is able to obtain novel, profession-specific perspectives beyond the predefined guide, whereas no emergent insights are extracted from \mimitalk, confirming our findings from \Cref{sec:auto_eval}. The moderate surprisal reflects that emergent insights build on familiar domain practices, as expected among expert participants. All emergent-insight differences are statistically significant ($p < 0.001$; \Cref{tab:sparkme_mimitalk}).

\begin{table*}[h]
\centering
\resizebox{\columnwidth}{!}{
\begin{tabular}{p{13.5cm} c c c c}
\hline
\textbf{Subtopic} & \multicolumn{2}{c}{\textbf{\systemname}} & \multicolumn{2}{c}{\textbf{\mimitalk}} \\
 & \textbf{Participant 1} & \textbf{Participant 2} & \textbf{Participant 3} & \textbf{Participant 4} \\
\hline
\multicolumn{5}{l}{\textbf{Predefined subtopics}} \\
\hline
General outlook on AI's broader societal and industry impact & \checkmark & \checkmark & - & \checkmark \\
Personal beliefs about the ethics and risks of AI in the workplace & \checkmark & \checkmark & \checkmark & - \\
Missing AI tools or features that would be most beneficial in the future & \checkmark & \checkmark & \checkmark & \checkmark \\
Predicted evolution of their job in the next 5--10 years with AI integration & \checkmark & \checkmark & \checkmark & \checkmark \\
Concrete steps they would want their organization to take regarding AI strategy & \checkmark & \checkmark & - & \checkmark \\
\hline
\multicolumn{5}{l}{\textbf{Emergent subtopics}} \\
\hline
Informal social norms and lack of ethical AI discussions among peers & \checkmark & - & - & - \\
Need for human oversight and ethical safeguards in sensitive AI use & \checkmark & - & - & - \\
Concrete organizational strategies for responsible AI adoption & \checkmark & - & - & - \\
Impact of emerging AI-specific roles on workflows and team dynamics & - & \checkmark & - & - \\
Impact of AI-related academic or organizational policies on workload & - & \checkmark & - & - \\
\hline
\end{tabular}}
\caption{An illustrative coverage comparison across 4 participants interviewed with \systemname and \mimitalk on the topic "AI Attitudes and Future Outlook". While both systems and participants follow the same interview topic guide, \systemname consistently covers predefined subtopics while eliciting emergent themes that differ across interviews. In contrast, \mimitalk covers a subset of predefined topics and does not surface additional themes.}
\label{tab:example-userstudy-comparison}
\end{table*}

\paragraph{Illustrative Example of Insights.} Table~\ref{tab:example-userstudy-comparison} illustrates the coverage comparison for $4$ different participants between \systemname and \mimitalk for the topic "AI Attitudes and Future Outlook." While both \systemname and \mimitalk cover subtopics in the topic guide,
\systemname also surfaces emergent subtopics that vary across participants: for one, themes around informal social norms and organizational safeguards arise; for another, themes around AI-specific roles and institutional policy changes emerge instead. This shows the value of \systemname in conducting a study. Consider a study in the formative stage, when a fully exhaustive guide might not be present.  The emergent subtopics from different interviews can be interpreted in two ways---recurring emergent subtopics across participants are candidates for addition to future versions of the guide, while those that are participant-specific can be examined case-by-case by researchers as desired. Additional examples are provided in \Cref{sec:app-example-high-qual-emerg}.
\section{Related Work}
\label{sec:related}

\paragraph{Formulation of the interview objective.}
\label{sec:related_objective}
Semi-structured interviews combine a predefined topic guide with adaptive follow-up based on participant responses, while requiring interviewers to avoid introducing assumptions that may shape participants' accounts~\citep{weiss1995learning, kvale2009interviews, adams2015conducting, cairns2022enhancing}. We formulate an objective for semi-structured interviewing (\Cref{sec:interview_formulation}) that is inspired by work on task-oriented dialogue. \citet{walker1997paradise} formalized dialogue evaluation as a weighted trade-off between task success and interaction cost that was operationalized in dialogue systems for various tasks \citep{levin1998using, lemon2006isu}. More recent work has made this trade-off explicit through multi-objective reinforcement learning formulations \citep{ultes2017reward, chu2023multi}. A complementary line of research has focused on learning to ask questions that maximize information gain under turn constraints, framing dialogue as an efficient information-seeking process \citep{chen2018learning, choi2018quac}. Our formulation builds on these ideas by adapting them to semi-structured interviewing, where task success is defined not only by coverage of predefined topics but also by the discovery and exploration of participant-introduced emergent subtopics, reflecting the distinctive goals of qualitative inquiry \citep{adams2015conducting, weiss1995learning, kvale2009interviews}.

\paragraph{Conducting interviews with LLM-based systems}
Early non-LLM systems explored automated interviewing using rule-based dialogue management and predefined probing logic \citep{xiao2020tell}. The improvement in LLM capabilities has led to a growing body of recent work that has explored automated systems for conducting interviews and interview-like interactions. One line of work relies on a \emph{single LLM} guided by carefully designed prompts to emulate human interviewers, conducting semi-structured or adaptive interviews and logging responses for downstream analysis \citep{liu2025understand, wuttke2025ai, liu2024step, wei2024leveraging, cuevas2025collecting, park2024generative, gomez2025virtual, handa2025interviewer, spangher2025newsinterview, geieckejaravel2026}.
These systems typically encode interviewing strategy through prompt design, question templates, or persona conditioning, and demonstrate the feasibility of using prompted LLMs for conducting interviews. \llmbaseline and \interviewgpt are representative baselines for this direction.
The second line of recent work uses multiple independently prompted LLM-agents that decompose interviewing into specialized roles, such as planning, question asking, response interpretation, and oversight, enabling more explicit coordination and control over the interview process \citep{talaei2025storysage, liu2025mimitalk, pathak2025ai}. \mimitalk and \storysage correspond to this direction.
To our knowledge, prior work does not explicitly formalize an interview objective that balances topic guide coverage, exploration of participant-introduced themes, and interview cost.

\section{Conclusion}
\label{sec:conclusion}

In this work, we introduce a tractable, customizable objective for semi-structured interviewing that formalizes the trade-off between covering a predefined interview topic guide, exploring emergent themes grounded in participant responses, and minimizing interview cost. Building on this formulation, we design \systemname, a multi-agent interviewer that incorporates an explicit planning step with simulated conversation rollouts to guide exploratory questioning. In a controlled benchmark with simulated user agents, we compare against prior interview systems and show that \systemname improves both coverage and emergent theme discovery. We further validate \systemname in a human user study with $70$ domain experts across $7$ professions, demonstrating that these improvements translate into the elicitation of relevant and peer-validated novel insights in real-world qualitative data collection, and qualitative-research experts find stronger responsiveness and probing in its interviews.

Our work opens multiple directions for future research. We release \systemname as an open-source system for researchers in different domains to customize and deploy at scale to support real-world human studies.
Our objective formulation also suggests a methodological direction for training LLM-based interviewers directly. While typical training objectives of LLMs focus on engagement and human preferences, our utility score can form the basis of training models that fulfill multi-turn objectives. As collecting human interview data at scale is expensive and challenging, the current bottleneck is the performance of user agents, which we detail in \Cref{sec:limitations}. As user simulation models improve, the proposed objective could be used as a training signal to learn interviewer policies through automatic rollouts. 

\section{Limitations and Future Work}
\label{sec:limitations}

We first note that in our user study, we do not compare \systemname directly against expert human interviewers. While our evaluation benchmarks against existing LLM-based systems, human interviewers remain the gold standard for semi-structured interviewing. The qualitative-research experts in our additional evaluation assess completed transcripts rather than conduct the interviews themselves. Prior to real-world deployments, we caution that systems such as \systemname should be carefully evaluated in the target domains against expert human interviews.

Second, we note that our automatic evaluation (\Cref{sec:auto_eval}) relies on LLM-based user agents and automatic metrics. While these have the advantage of being controlled and repeatable, allowing comparisons between interview systems, we note that these user agents may respond differently from human participants and cannot fully capture the diversity of human behavior~\citep{zhou2026mind,vijayvargiya2026lost}. We present these results for structured benchmarking and ablating the various aspects of our system design and perform extensive user evaluation to confirm that \systemname yields relevant and novel insights in \Cref{sec:user_study}. We also note that it is not possible to try out all the variants of prompts that might elicit differing outcomes for the baselines. We restrict ourselves to prompts released by the various baseline works and share the exact implementation used for the results in this paper.

Finally, our automatic evaluation relies on LLM-as-a-judge metrics to enable controlled comparison across systems. We validate the predefined-coverage measure against human judgments (\Cref{sec:coverage_llm_as_judge}), but this validation may not generalize to other domains or interview settings.

\section*{Acknowledgments}

We would like to thank members of the Stanford NLP group, the SALT Lab, and the Stanford Digital Economy Lab for their helpful discussions at different stages of the project. We thank the participants for the user study recruited from Upwork. We also greatly benefited from feedback on the written draft from Chenglei Si, Hao Zhu, Yutong Zhang, Joachim Baumann, Jennifer Wang, Ryan Louie, Yijia Shao, Luca Vendraminelli, Joseph Chee Chang, Matthew Ordrick Sanjaya, Yuhao Yang, and Chenyue Li. This work was supported by an HAI grant, DSO lab, Open Philanthropy, Schmidt Sciences, a grant under the NSF CAREER IIS-2247357 and ONR N00014-24-1-2532, and support from the Stanford Digital Economy Lab.

\bibliographystyle{plainnat}
\bibliography{custom}

\appendix
\newpage

\section{Additional Experimental Details and Results}
\label{sec:appendix_expt_details}

\subsection{Additional Automatic Evaluation Results}
\label{sec:appendix_results}

\begin{figure*}[!t]
\centering
    \includegraphics[width=\textwidth]{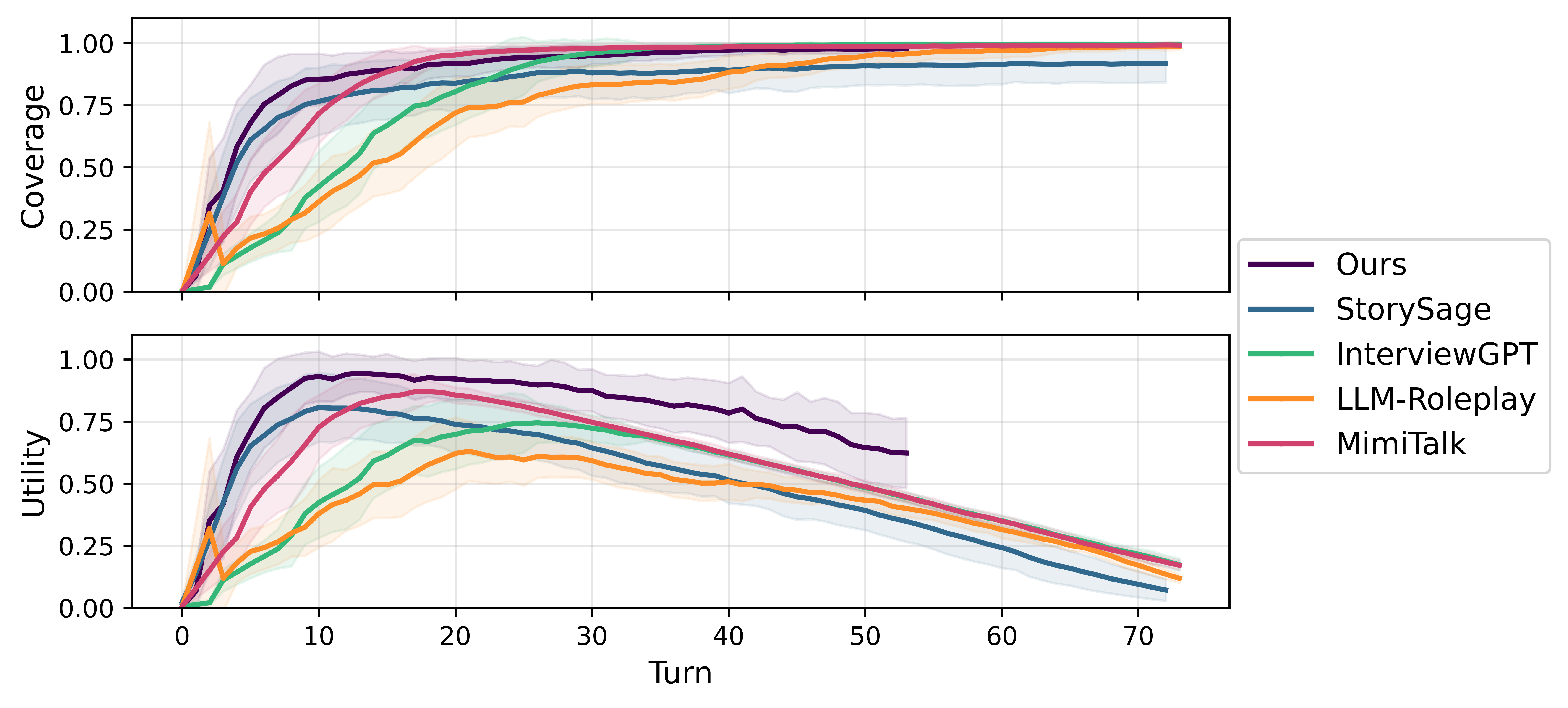}
    \caption{(a) Predefined subtopic coverage and (b) utility of subtopics ($y$-axis) as a function of the number of interview turns ($x$-axis) for different systems (\Cref{sec:baselines}), using \texttt{GPT-4.1-mini} as the backbone for the interviewer LLM and user agent. \systemname (Ours) efficiently converges to a higher coverage and utility value, consistently outperforming other baselines}
    \label{fig:coverage-utility-comparison-gpt41mini}
\end{figure*}

\paragraph{We observe similar trends with different interviewer models.} Figure~\ref{fig:coverage-utility-comparison-gpt41mini} shows that \systemname consistently outperforms all four baselines in overall interview utility and is competitive in terms of coverage. In contrast with using \texttt{Qwen3-30B-A3B-Instruct}, \texttt{GPT-4.1-mini} prioritizes coverage first, but then later there is a sudden shift toward focusing more on emergence. To be specific, \systemname achieves a better interview utility score, with the highest peak utility score ($0.943$), reaching this peak as early as turn $13$, which reflects effective coverage of predefined topics while also being able to gather emergent content during the interview. Among the baselines, \mimitalk performs strongest ($0.870$), followed by \interviewgpt ($0.744$), with \storysage ($0.806$) and \llmbaseline ($0.630$) trailing further behind. In terms of coverage, \systemname achieves an average coverage score of $0.977$ within at most $53$ turns, outperforming \storysage ($0.918$), while still competitive with \interviewgpt ($0.995$), \llmbaseline ($0.988$), and \mimitalk ($0.992$).

\paragraph{Adjusting the internal weighting of the objectives changes the amount of exploration in \systemname interviews.} 

\begin{figure*}[!t]
\centering
    \includegraphics[width=\textwidth]{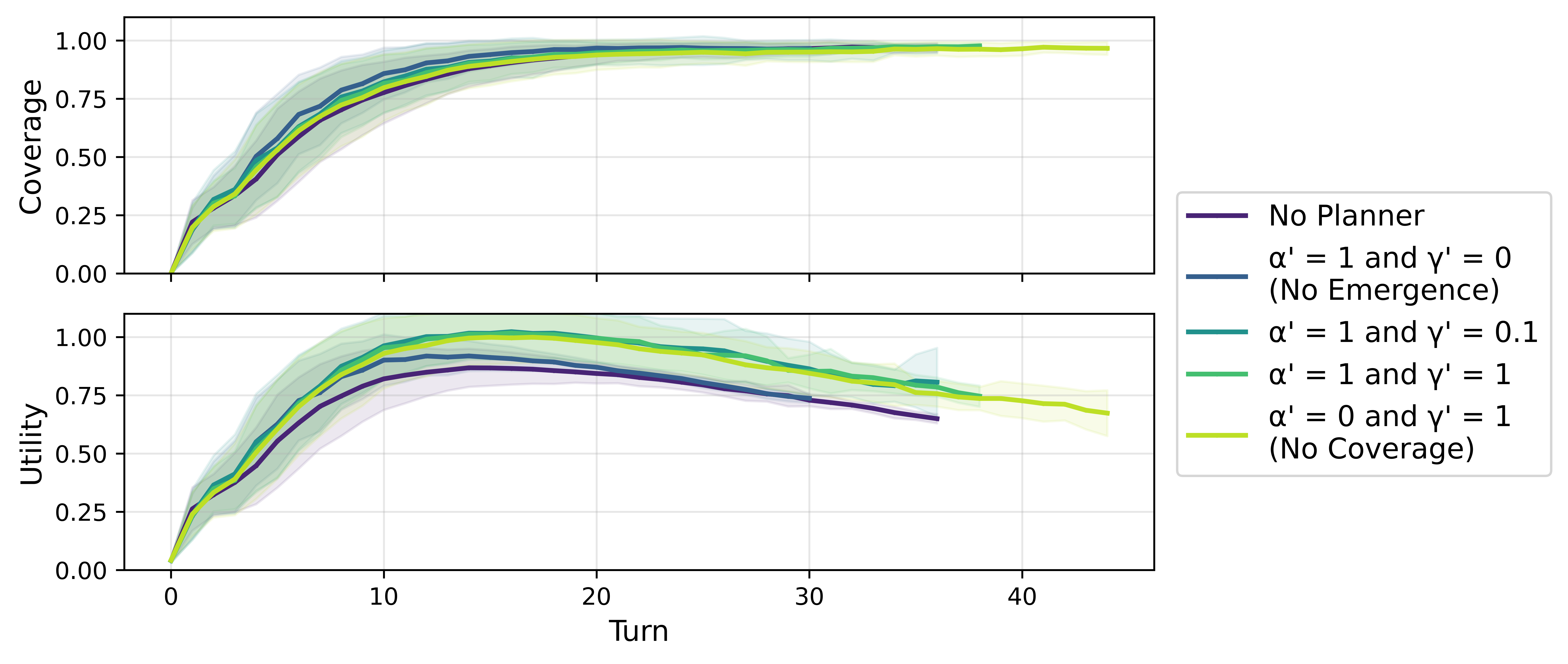}
    \caption{Predefined subtopic coverage and utility of subtopics ($y$-axis) as a function of the number of interview turns ($x$-axis) by ablating different features of \systemname, using \texttt{Qwen3-30B-A3B-Instruct-2507} as the backbone for the interviewer LLM and user agent.
    }
    \label{fig:coverage-utility-comparison}
\end{figure*}

To demonstrate the customizability of our system, we report results by varying the internal weights assigned to the predefined topic guide coverage ($\alpha'$) and emergent subtopics ($\gamma'$) within the \ep (\Cref{sec:coverage-planner}). As a result, the directions prioritized during the simulated rollouts change. We also compare the full system to \systemname without \ep. The predefined subtopic coverage and utility of these ablations are shown in \Cref{fig:coverage-utility-comparison}.\footnote{Note that in the evaluation of utility in \Cref{fig:coverage-utility-comparison}, we still set the weight of $\alpha = \frac{1}{48}$, $\beta = \frac{1}{72}$, and $\gamma = \frac{1}{24}$ uniformly to demonstrate the different outcomes between the baselines.} We find that setting the internal weight of $\gamma$ to zero leads to a lower utility value (\Cref{fig:coverage-utility-comparison}), due to covering fewer emergent subtopics (\Cref{fig:emergence-box}). This is comparable to the baseline without \ep. However, in these cases, we note that \systemname covers the predefined subtopics in the topic guide more efficiently. Depending on the desired outcome, researchers can use \systemname appropriately. We do note that setting a weight of zero to $\alpha'$, or predefined subtopic coverage, in \ep still leads to interviews that cover the topic guide as \ia and \am keep the conversation on track. 
In practice, these weights provide a direct mechanism for researchers to control the interview style according to their needs. We find that meaningful behavioral differences emerge primarily when there is a sufficiently large ratio contrast between $\alpha'$ and $\gamma'$ rather than from fine-grained tuning of exact values. For instance, in a formative qualitative study, we might set a higher $\gamma'$ value to encourage exploratory interviews that prioritize uncovering emergent themes. In contrast, in a structured setting, such as quantifying user opinions or sentiment on particular topics, where we want time-efficient interviews that strictly adhere to the topic guide, we would set a higher $\alpha'$ value. We recommend selecting weights based on the study objective.

\begin{figure*}[!t]
\centering
    \includegraphics[width=\textwidth]{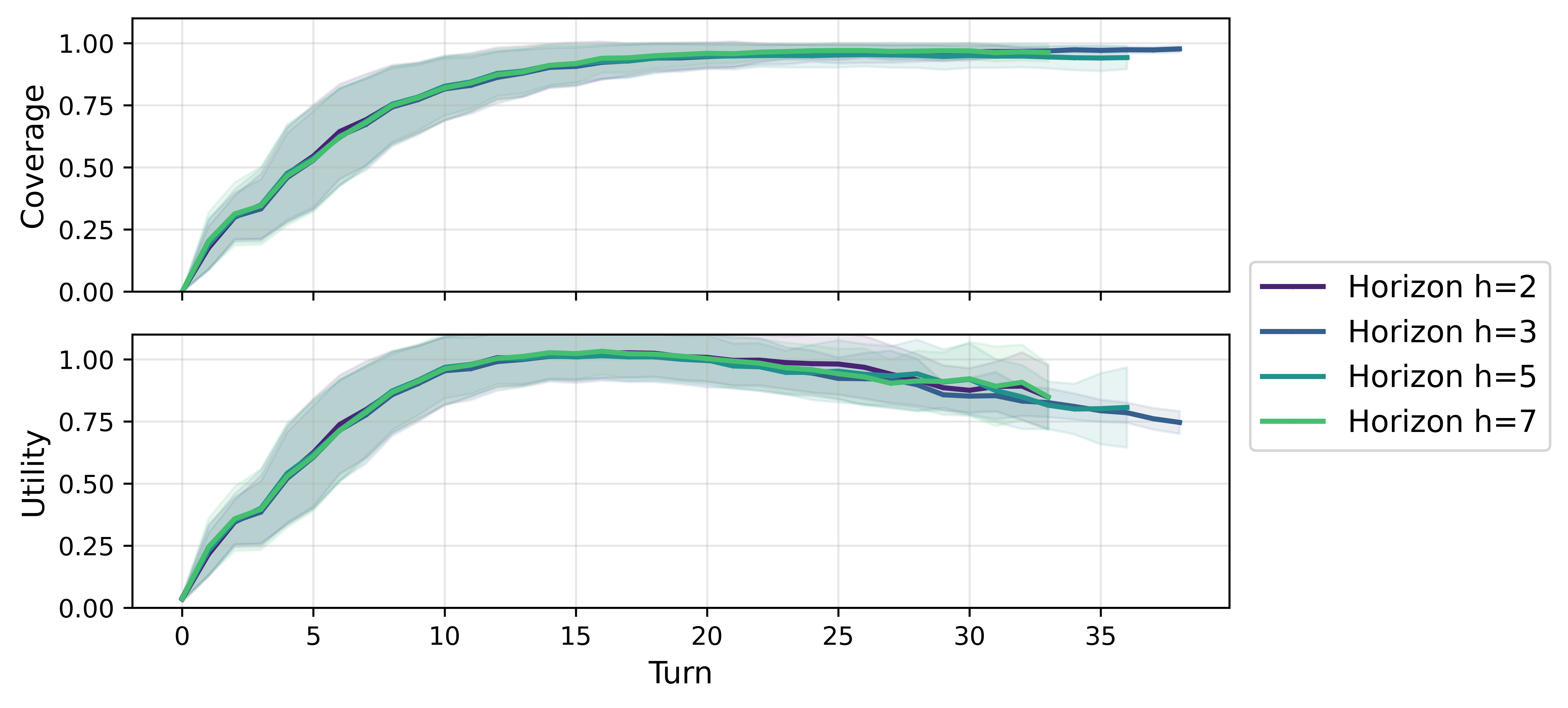}
    \caption{(a) Predefined subtopic coverage and (b) utility of subtopics ($y$-axis) as a function of the number of interview turns ($x$-axis) for \systemname, by varying the horizon length, using \texttt{Qwen3-30B-A3B-Instruct-2507} as the backbone for the interviewer LLM and user agent.
    }
    \label{fig:coverage-utility-comparison-horizon}
\end{figure*}

\paragraph{Increasing the horizon of rollouts does not lead to higher utility.} In \Cref{fig:coverage-utility-comparison-horizon}, we report the predefined subtopic coverage and utility when increasing the horizon of rollouts performed by \ep. We find that this leads to minimal difference in performance, explained by the observation that the interviewer tends to sufficiently cover a subtopic in 1--2 questions, so estimating the progress of the conversation beyond that point tends to lead to noisy rollouts and limited additional signal.

\begin{figure*}[!t]
\centering
    \includegraphics[width=\textwidth]{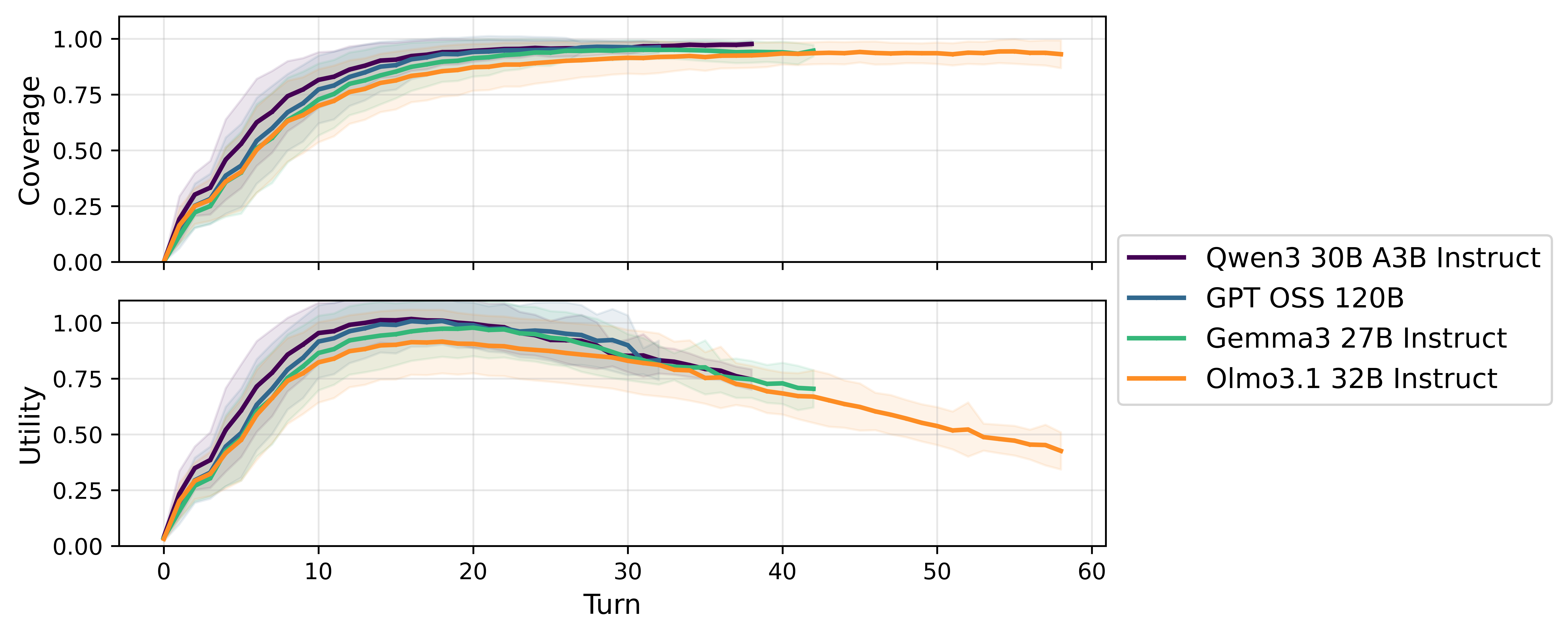}
    \caption{(a) Predefined subtopic coverage and (b) utility of subtopics ($y$-axis) as a function of the number of interview turns ($x$-axis) for \systemname using \texttt{Qwen3-30B-A3B-Instruct-2507} as the backbone for the interviewer LLM and various LLMs for the user agent.
    }
    \label{fig:coverage-utility-comparison-diff-user}
\end{figure*}

\paragraph{Varying the interviewee user agent model leads to largely similar trends.} We note that the results in \Cref{sec:auto_eval} use the same model for the interviewer system and interviewee user agent. To confirm that this is not model-specific behavior, we vary the user agent model and report results on predefined subtopic coverage and utility in \Cref{fig:coverage-utility-comparison-diff-user}. The interviewer is \texttt{Qwen-3-30B-A3B-Instruct} for each of these plots. The trends tend to largely hold for the different models with faster convergence to a utility value comparable to \texttt{Qwen-3-30B-A3B-Instruct} for the stronger model, \texttt{GPT-OSS-120B}. The weaker models, \texttt{OLMo-3.1-32B-Instruct} and \texttt{Gemma-3-27-Instruct}, converge to marginally lower values of both predefined topic coverage and utility. Overall, these results provide additional evidence that our conclusions about the design of \systemname from \Cref{sec:auto_eval} are robust to model choice.

\begin{figure*}[!t]
    \centering
    \includegraphics[width=.85\textwidth]{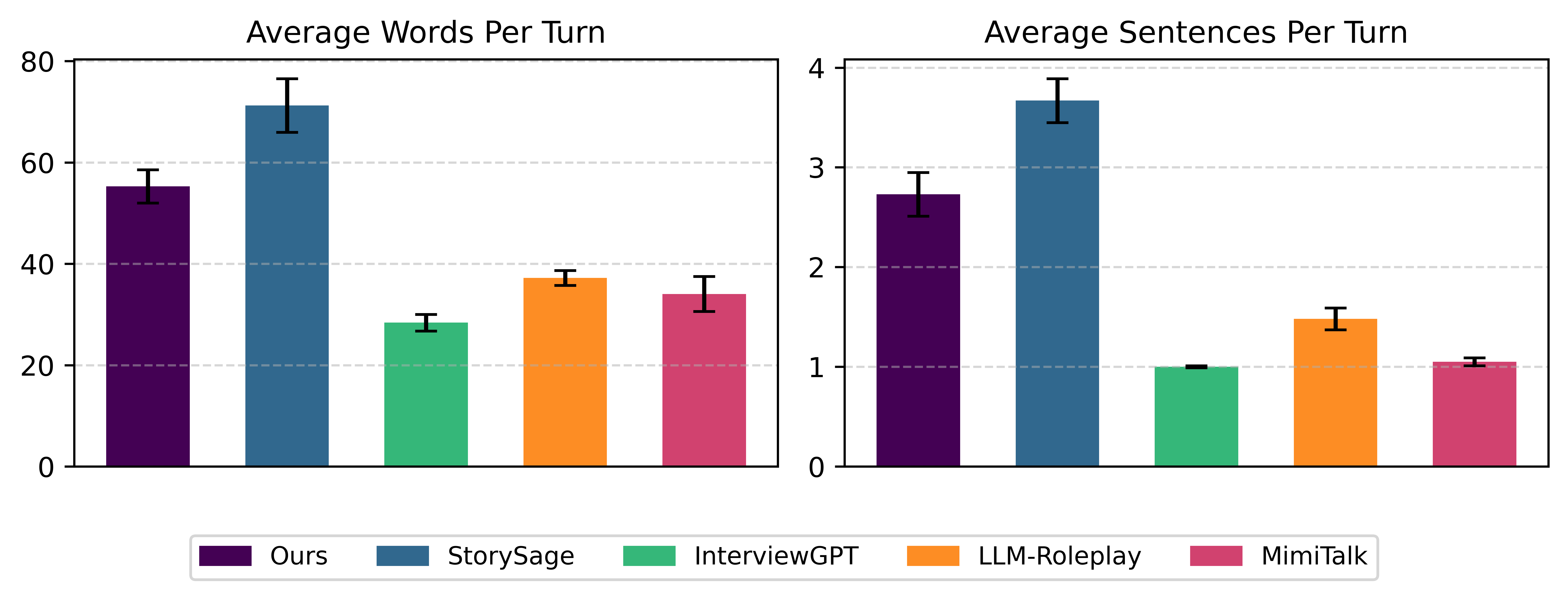}
    \caption{Statistics per turn between \systemname, \llmbaseline, \storysage, \mimitalk, and \interviewgpt.}
    \label{fig:statswords}
\end{figure*}

\begin{figure*}[!t]
    \centering
    \includegraphics[width=.85\textwidth]{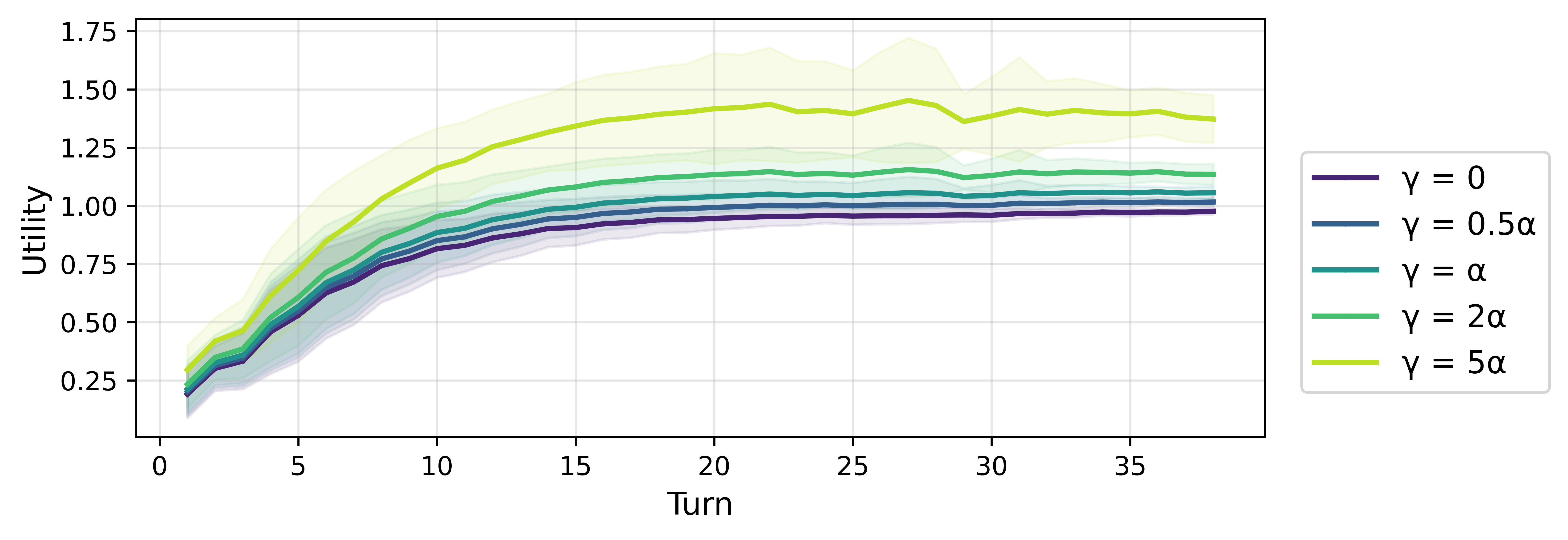}
    \caption{Utility score for different $\gamma$ with $\alpha = \frac{1}{48}$ and $\beta$ of 0.}
    \label{fig:interview-gamma-vary}
\end{figure*}

\paragraph{Additional auxiliary statistics.}
Figure~\ref{fig:statswords} shows the average number of words per turn and the average number of sentences per turn for each system. Figure~\ref{fig:interview-gamma-vary} illustrates the effect of varying $\gamma$ while fixing $\alpha = \frac{1}{48}$ and $\beta = 0$ for evaluation.

\subsection{Additional User Study Results}
\label{sec:appendix_user_study_results}

\begin{table*}[t]
\centering
\small
\begin{tabular}{l cc ccc}
\toprule
\textbf{Metric} & \textbf{SparkMe} & \textbf{MimiTalk} & $t$ & $p$ & Cohen's $d$ \\
\midrule
\multicolumn{6}{l}{\textit{Content Quality}} \\
\quad Coverage              & $4.37 \pm 0.31$ & $3.14 \pm 0.81$ & 8.39  & $1.17 \times 10^{-10}$ & $2.01^{***}$ \\
\quad Depth                 & $4.06 \pm 0.62$ & $3.09 \pm 0.85$ & 5.45  & $9.03 \times 10^{-7}$  & $1.30^{***}$ \\
\quad Correctness           & $4.23 \pm 0.45$ & $3.46 \pm 0.92$ & 4.45  & $4.93 \times 10^{-5}$  & $1.06^{***}$ \\
\midrule
\multicolumn{6}{l}{\textit{Interaction Quality}} \\
\quad Clarity               & $4.14 \pm 0.30$ & $3.89 \pm 0.96$ & 1.47  & $0.149$                & $0.35$ \\
\quad Adaptiveness          & $4.09 \pm 0.41$ & $3.49 \pm 0.92$ & 3.52  & $9.59 \times 10^{-4}$  & $0.84^{***}$ \\
\quad Comfort               & $4.11 \pm 0.25$ & $4.09 \pm 0.89$ & 0.13  & $0.899$                & $0.03$ \\
\midrule
\multicolumn{6}{l}{\textit{Emergent Insights}} \\
\quad Topic Relevance       & $4.14 \pm 0.50$ & $1.00 \pm 0.00$ & 37.15 & $< 10^{-16}$           & $8.88^{***}$ \\
\quad Analytical Value      & $3.50 \pm 0.47$ & $1.00 \pm 0.00$ & 31.47 & $< 10^{-16}$           & $7.52^{***}$ \\
\quad Topic-level Emergence & $3.34 \pm 0.45$ & $1.00 \pm 0.00$ & 30.76 & $< 10^{-16}$           & $7.35^{***}$ \\
\quad Surprise              & $2.62 \pm 0.36$ & $1.00 \pm 0.00$ & 26.62 & $< 10^{-16}$           & $6.36^{***}$ \\
\midrule
\multicolumn{6}{l}{\textit{Overall}} \\
\quad Experience            & $4.26 \pm 0.28$ & $3.49 \pm 0.89$ & 4.88  & $1.66 \times 10^{-5}$  & $1.17^{***}$ \\
\bottomrule
\end{tabular}
\caption{Welch's $t$-test comparing \systemname and \mimitalk based on human study ratings ($n=35$ per group). $^{***}$ indicates $p < 0.001$.}
\label{tab:sparkme_mimitalk}
\end{table*}

\paragraph{Full user study results.} Table~\ref{tab:sparkme_mimitalk} reports more detailed numbers for the user study results, along with the complete Welch's $t$-test results. \systemname significantly outperforms \mimitalk ($p<0.001$) on all content quality metrics (coverage, depth, correctness), on adaptiveness, on all emergent insight metrics (topic relevance, analytical value, topic-level emergence, surprise), and on overall experience. On the other hand, the differences in clarity ($p=0.149$) and comfort ($p=0.899$) are not significant.

\subsubsection{Evaluation by Qualitative and Human-Centered Research Experts}
\label{app:sec-hci-expert-eval-results}

To complement validation by occupational domain experts, we conduct a transcript-based evaluation with five experts who have PhDs and relevant experience in qualitative research, UX, or HCI. We recruit the evaluators through Upwork and compensate them at \$30/hour. These experts evaluate 20 de-identified transcripts from the user study (10 from \systemname and 10 from \mimitalk), which are organized into 10 profession-matched pairs. Each transcript receives two independent ratings, with both transcripts in a matched pair assigned to the same two evaluators. Each evaluator reviews eight transcripts, four from each system, before providing a written comparison of the two anonymized transcript sets.

The expert evaluators assess the following using five-point scales: (1) \emph{preservation of participant authorship}, or whether questions minimized interviewer-imposed assumptions and framing; (2) \emph{contingent responsiveness}, or whether follow-ups built accurately on participants' responses; and (3) \emph{probing quality}, or whether follow-ups elicited useful detail, examples, clarification, and explanation. The complete instructions and rating rubrics can be found in~\Cref{sec:app-hci-eval-rubrics}.

We average the two ratings for each transcript and compare the systems within profession-matched pairs using two-sided exact sign-flip tests. We apply Holm correction across the three dimensions and compute 95\% percentile confidence intervals from 100,000 matched-pair bootstrap resamples. We report ordinal Krippendorff's $\alpha$ to quantify inter-evaluator agreement.

\begin{table*}[t]
    \centering
    \small
    \setlength{\tabcolsep}{3.5pt}
    \begin{tabular}{lccccc}
        \toprule
        Dimension & \systemname & \mimitalk & Difference [95\% CI] & $p_{\mathrm{Holm}}$ & $\alpha$ \\
        \midrule
        Participant authorship & $3.75\,(0.79)$ & $4.20\,(0.54)$ & $-0.45\,[-1.05,\,0.10]$ & $0.266$ & $0.483$ \\
        Contingent responsiveness & $4.00\,(0.58)$ & $2.95\,(0.96)$ & $1.05\,[0.45,\,1.60]$ & $0.047$ & $0.123$ \\
        Probing quality & $3.85\,(0.58)$ & $2.70\,(0.82)$ & $1.15\,[0.50,\,1.70]$ & $0.047$ & $0.148$ \\
        \bottomrule
    \end{tabular}
    \caption{Evaluation of interview quality by qualitative and human-centered research experts. Means and standard deviations (in parentheses) are computed over the 10 transcript-level means per system after averaging the two ratings for each transcript. Differences compare profession-matched transcripts. Confidence intervals use matched-pair bootstrap resampling; $p$-values are from exact paired sign-flip tests with Holm correction across the three dimensions. Higher scores are better.}
    \label{tab:hci-expert-interview-quality}
\end{table*}

As shown in Table~\ref{tab:hci-expert-interview-quality}, \systemname receives higher ratings than \mimitalk for contingent responsiveness ($4.00$ vs.\ $2.95$) and probing quality ($3.85$ vs.\ $2.70$). Both differences remain significant after Holm correction ($p_{\mathrm{Holm}}=0.047$), and \systemname is favored in eight of the 10 matched pairs on each dimension. While \mimitalk receives a numerically higher rating for preservation of participant authorship ($4.20$ vs.\ $3.75$), this difference is not significant ($p_{\mathrm{Holm}}=0.266$). This interview-level human assessment is consistent with the broader automated question-level audit, which finds severe leading questions to be rare for both systems (\Cref{sec:leading-question-eval}).

The absolute-score agreement varies across dimensions, with ordinal Krippendorff's $\alpha$ of $0.483$ for participant authorship, $0.123$ for responsiveness, and $0.148$ for probing. Nevertheless, the direction of the system comparison is more consistent: all five evaluators favor \systemname for responsiveness at the condition level, and four of five favor it for probing. This suggests that evaluators differ in how they calibrate individual Likert scores but generally identify the same comparative strengths.

The final written comparisons from the evaluators also support this interpretation. Evaluators characterize \systemname as more responsive to participant-provided details and more likely to request concrete examples, explanations, and consequences. In contrast, \mimitalk is more frequently described as following a predetermined sequence, repeating previously covered material, or moving past responses that could be explored further. Evaluators also mention that \mimitalk generally asks shorter and more neutral questions, whereas \systemname sometimes uses verbose, multi-part, or highly framed questions that may restrict participant authorship. Thus, the evaluation supports a specific advantage in adaptive follow-up and probing rather than an unqualified improvement across all aspects of interview quality.

\subsubsection{Automated Leading-Question Audit}
\label{sec:leading-question-eval}

In addition to collecting user ratings, we evaluate the presence and use of leading questions in the user study interviews as a measure of data quality and interpretive validity. Leading questions are questions that steer participants' responses by introducing new content, embedding presuppositions, or implicitly evaluating an answer, which could influence qualitative responses and threaten data reliability~\citep{loftus1975leading, swann1982leading, poole1995interviewing, cairns2022enhancing}. We use the cleanness framework introduced by~\citet{cairns2022enhancing}, which follows the principles of clean questioning by~\citet{tosey2014eliciting}. The instruction prompt and cleanness rubric are provided in Figure~\ref{prompt:leading-question-prompt}. Because the criteria for scores of $3$, $4$, and $5$ are highly similar, we collapse scores $3$, $4$, and $5$ into a single category for this analysis, resulting in a three-point scale.

We use \texttt{mR3-Qwen3-14B}~\citep{anugraha2025mr3}, a rubric-based reward model, as the evaluator. Overall, \mimitalk achieves an average score of $2.973 \pm 0.042$, while \systemname achieves $2.958 \pm 0.047$. Both systems therefore receive high average cleanness scores. The automated audit assigns a score of $1$, its lowest cleanness category, to two questions from \systemname and three from \mimitalk (Table~\ref{tab:all-leading-questions}).

One illustrative example from \systemname is:
\emph{``It's interesting that prompt writing has become an important skill for you. Could you tell me more about how learning to write effective prompts has changed the way you work with AI? Are there particular challenges or benefits you've noticed as you've developed this skill?''} While this question contains mild implicit guidance, it remains largely open-ended and does not strongly bias the participant's response.

\begin{table*}[t]
\centering
\begin{tabularx}{\textwidth}{lX}
\toprule
\textbf{System} & \textbf{Leading Questions} \\
\midrule
\systemname &
\textbf{Q1.} It's interesting that prompt writing has become an important skill for you. Could you tell me more about how learning to write effective prompts has changed the way you work with AI? Are there particular challenges or benefits you've noticed as you've developed this skill? \\ 
& \textbf{Q2.} It's great to hear you've had positive experiences with several AI platforms. Shifting gears a bit, I'm curious about how you feel AI has impacted the importance of your existing skills as a data scientist. Do you find that AI has enhanced the value of your current expertise, or has it reduced the need for some skills? How has that influenced the way you approach your work? \\
\midrule
\mimitalk &
\textbf{Q1.} What motivates you to dedicate 3 to 4 hours daily to learning new technologies, and are there any challenges you face in maintaining this learning habit? \\
& \textbf{Q2.} Can you tell me about a recent time when an AI chat tool helped you solve a tricky coding or IT issue? What was the problem and how did the AI assist you? \\
& \textbf{Q3.} Why do you feel that none of your tasks are suitable for automation? Are there specific reasons or concerns that make automation less appropriate? \\
\bottomrule
\end{tabularx}
\caption{Questions assigned the lowest cleanness score by the automated audit for \systemname and \mimitalk.}
\label{tab:all-leading-questions}
\end{table*}

\subsection{Examples of High-Quality Emergent Insights.}
\label{sec:app-example-high-qual-emerg}

To illustrate the value of emergent content surfaced by \systemname, we provide three emergent subtopics that received high peer ratings for both topic-level emergence and analytical value.\footnote{Due to the relatively small sample size, we caution against drawing takeaways about the workforce from these findings; we present these as a demonstration of the value of \systemname.}

\paragraph{Educators.} In our survey, educators highlighted an interesting case of AI integration in lesson planning workflows. One participant described using ChatGPT extensively to generate lesson content and activities, integrating outputs with Canva and Google Slides to structure daily presentations. They also noted that formatting inconsistencies often required manual adjustments. Peer annotators rated this insight highly, recognizing its practical value, while also noting that the manual adjustments demand significant time and effort, both indicating potential benefits of AI-assisted workflows and the ongoing challenges of human-AI collaboration in teaching.

\paragraph{HR Professionals.} Among HR and Administration professionals, one participant highlighted the growing importance of multilingual and in-person support skills in AI-augmented client services. They explained that fluency in the client’s language and demonstrating empathy were critical for overcoming communication barriers and reducing client anxiety. The participant noted that as AI automates some of their routine tasks, they can now focus more on building stronger interpersonal relationships, including practicing languages with AI assistance.

\paragraph{Creative Professionals.} Creative and Design professionals also shared perspectives on the future outlook for AI, particularly regarding copyright and legal constraints. One participant explained that these rules restrict AI usage to internal concept development, preventing AI-generated outputs from being used in final commercial work. As a result, teams must rely on traditional methods to complete artwork, which limits experimentation and slows learning about AI capabilities. Peer annotators agreed that this insight reflects important considerations for the future of AI in creative work, highlighting the need to balance innovation with legal and ethical constraints.

\subsection{Interview Topics and Subtopics}
\label{sec:app-interview-guide-list}

Table~\ref{tab:interview-topics-part1} and Table~\ref{tab:interview-topics-part2} show the topics and subtopics used for all of our interviews, which would be part of our interview topic guide-list.

\begin{table*}[!th]
    \centering
    \renewcommand{\arraystretch}{1.2}
    \begin{tabularx}{\textwidth}{@{} p{4cm} X @{}}
        \toprule
        \textbf{Topic} & \textbf{Subtopics} \\
        \midrule
        
        \textbf{Introduction \& Background} & 
        \begin{itemize}
            \item Educational background or training
            \item Specific job title and role description
            \item Current industry or sector (e.g., tech, finance, manufacturing)
            \item Company size and environment
            \item Type of business or market segment
            \item Duration/years of experience in current role
            \item Professional seniority or career level
        \end{itemize} \\
        \midrule

        \textbf{Core Responsibilities and Decision-Making} & 
        \begin{itemize}
            \item Primary job responsibilities and regular daily tasks
            \item Approximate proportion of time spent on core activities
            \item Level of autonomy and scope of decision-making in the role
        \end{itemize} \\
        \midrule

        \textbf{Task Proficiency, Challenge, and Engagement} & 
        \begin{itemize}
            \item Tasks that feel easiest or most natural to perform
            \item Tasks perceived as most challenging or complex
            \item Tasks that are repetitive, data-heavy, or suitable for automation
            \item Tasks that are most enjoyable or engaging versus those that feel boring or tedious
            \item Common pain points or inefficiencies in completing tasks
            \item How enjoyment, skill level, and productivity relate to one another
        \end{itemize} \\
        \midrule

        \textbf{Tech Learning Comfort} & 
        \begin{itemize}
            \item Attitude towards learning new technologies and tools
            \item Perceived adaptability to new software/methods
            \item Willingness to invest time in tech training
            \item Motivations or barriers to learning new tech (e.g., workload, relevance)
            \item Influence of peers or management on willingness to adopt new tools
        \end{itemize} \\
        \midrule

        \textbf{Primary Tools and Technologies Used in Work} & 
        \begin{itemize}
            \item Specific software, platforms, or systems used daily
            \item Essential non-AI tools for workflow
            \item Familiarity with industry-standard technologies
            \item Interoperability or integration issues between tools
        \end{itemize} \\
        \bottomrule
    \end{tabularx}
    \caption{Interview Topics and Subtopics (Part 1 of 2)}
    \label{tab:interview-topics-part1}
\end{table*}

\begin{table*}[!th]
    \centering
    \renewcommand{\arraystretch}{1.2}
    \begin{tabularx}{\textwidth}{@{} p{4cm} X @{}}
        \toprule
        \textbf{Topic} & \textbf{Subtopics} \\
        \midrule

        \textbf{AI Experience and Tool Adoption} & 
        \begin{itemize}
            \item Familiarity with fundamental AI/ML concepts and terminology
            \item Names of specific AI software/platforms currently used in work
            \item Frequency and purpose of AI tool application (specific use cases)
            \item Specific examples of AI success and failure experiences (lessons learned)
            \item Availability of organizational training or peer resources for AI use
        \end{itemize} \\
        \midrule

        \textbf{AI Interaction Style and Workflow Change} & 
        \begin{itemize}
            \item Preferred mode of interaction: independent versus step-by-step collaboration
            \item Style of human-AI teaming (e.g., advisor, assistant, co-worker)
            \item Willingness and openness to adopting new AI-driven workflows
            \item Preference for conversational vs. command-based interfaces (communication dynamics)
        \end{itemize} \\
        \midrule

        \textbf{Trust and Control Over AI} & 
        \begin{itemize}
            \item Extent to which tasks rely on specialized, tacit domain knowledge
            \item Level of trust in AI outputs for work tasks and critical decisions
            \item Ideal balance of human effort and AI automation for specific tasks
            \item Conditions under which high automation is acceptable or threatening
        \end{itemize} \\
        \midrule

        \textbf{AI Impact on Skills and Job Security} & 
        \begin{itemize}
            \item Perceived impact of AI on the importance of existing skills (enhanced vs. reduced)
            \item Emerging skills or new areas of responsibility created by AI
            \item Level of concern about AI replacing specific tasks or the overall role
            \item Availability of override mechanisms or manual checks for AI-driven processes
            \item Perceived change in team or company policies regarding AI adoption
        \end{itemize} \\
        \midrule

        \textbf{AI Attitudes and Future Outlook} & 
        \begin{itemize}
            \item General outlook on AI's broader societal and industry impact
            \item Personal beliefs about the ethics and risks of AI in the workplace
            \item Missing AI tools or features that would be most beneficial in the future
            \item Predicted evolution of their job in the next 5-10 years with AI integration
            \item Concrete steps they would want their organization to take regarding AI strategy
        \end{itemize} \\
        \bottomrule
    \end{tabularx}
 \caption{Interview Topics and Subtopics (Part 2 of 2)}
 \label{tab:interview-topics-part2}
\end{table*}

\subsection{Examples of emergent subtopics in automatic evaluation}
\label{sec:examples_auto_eval_emergence}
An example of an emergent subtopic identified by \systemname is the impact of AI-generated content on local market nuance and cultural context in property descriptions. The interviewee noted that AI-generated property descriptions frequently lack local market characteristics and culturally specific context, necessitating manual regional review to ensure accuracy.

Another emergent subtopic concerns automated real-time regulatory compliance validation in design workflows. The interviewee expressed a desire for an AI tool that can automatically validate engineering designs against live FAA and EASA regulatory standards during the design phase, in order to eliminate the inefficiencies and risks associated with manual cross-referencing. Upon further probing, this need was grounded in a recent project delay caused by a missed EASA regulatory update.

\subsection{Sample user profile for automatic evaluation}
\label{sec:app_user_profile}

Figure~\ref{prompt:sample-user-profile} shows a snippet of an example of a synthetic user profile derived from one of the WorkBank workers~\citep{shao2025future}.

\begin{promptbox}[label=prompt:sample-user-profile]{A snippet of a user profile}
\begin{Verbatim}[
    fontsize=\footnotesize,
    breaklines=true,
    breakanywhere=true,
    breaksymbol=,
    breaksymbolleft=,
    breaksymbolright=,
    commandchars=\\\{\}
]
Topic: Introduction & Background

Subtopic ID: 1.1
Subtopic Description: Educational background or training
Notes:
- Holds an Associate Degree from a community college
- Studied a field related to industrial technology or manufacturing processes
- Completed coursework in technical mathematics, blueprint reading, and manufacturing safety
- Participated in hands-on lab work with CNC machinery during education

Subtopic ID: 1.2
Subtopic Description: Specific job title and role description
Notes:
- Works as a Production, Planning, and Expediting Clerk
- Responsible for coordinating production schedules and expediting the flow of materials
- Manages daily tasks involving assembly, CNC programming, and product shipment preparation
- Builds timers and alarms, ensuring all parts are manufactured and assembled to specification

...
\end{Verbatim}
\end{promptbox}

\subsection{Prompt for User Agent}

We simulate the User Agent as the interviewee using the prompt in Figure~\ref{prompt:user-agent-template}.

\begin{promptbox}[label=prompt:user-agent-template]{UserAgent Templates}
\begin{Verbatim}[
    fontsize=\footnotesize,
    breaklines=true,
    breakanywhere=true,
    breaksymbol=,
    breaksymbolleft=,
    breaksymbolright=,
    commandchars=\\\{\}
]
RESPOND_TO_QUESTION_PROMPT = """
{CONTEXT}

{PROFILE_BACKGROUND}

{CHAT_HISTORY}

{INSTRUCTIONS}

{OUTPUT_FORMAT}
"""

RESPOND_CONTEXT = """
<context>
You are playing the role of a real person being interviewed. You are currently in an interview session.

You now need to respond: provide a natural response that aligns with your character's personality and background, as if you are having a genuine conversation with an interviewer.
If this is the first turn, you should only say that you are happy to start the interview.
</context>
"""

PROFILE_BACKGROUND_PROMPT = """
This is your background information.
<profile_background>
{profile_background}
</profile_background>

Here are summaries from your previous interview sessions:
<session_history>
{session_history}
</session_history>
"""

CHAT_HISTORY = """
Here is the conversation history of your interview session so far. You are the <UserAgent>  in the chat history and you need to respond to the interviewer's last question.
<chat_history>
{chat_history}
</chat_history>
"""

RESPOND_INSTRUCTIONS_PROMPT = """
<instructions>
# GENERAL INTERVIEW RULES
- Always answer the question asked.
- Never skip a question.
- Do not anticipate follow-up questions.
- Treat this as a real interview: the interviewer controls depth and direction.
- Answer only what is necessary for the current question.

# HUMAN STOPPING HEURISTIC (CRITICAL)
Humans stop talking once they have given a sufficient answer, not a complete one.

- Aim for the first reasonable stopping point.
- Assume the interviewer may interrupt or follow up.
- Do not try to close the topic yourself.

# BREVITY & DEPTH CONTROL (STRICT)
- Answer length: 1-2 sentences.
- Typical answers should reveal approximately one concrete fact or signal.
- Do not compress multiple ideas, timelines, or facts into one answer.

# VAGUE OR OPEN-ENDED QUESTIONS (CRITICAL)
If a question is vague, ambiguous, or open-ended to answer without guessing, for example if it sounds like listing some list of topics, then:
- Do NOT invent scope or details.
- Briefly acknowledge the ambiguity (e.g., "I'm not sure which aspect you mean").
- Either ask one short clarification question, OR state one reasonable assumption and answer briefly under that assumption.
- Do not do both and do not expand beyond the assumed or clarified scope.

# CONTENT GUIDELINES
- Stay tightly focused on the question’s scope.
- Do not expand across time, roles, or institutions unless asked.
- Do not repeat prior answers unless explicitly prompted.
- Avoid lists unless the interviewer asks for them.
- Avoid meta-commentary about motivation, passion, or energy.

# EMERGENCE (ALLOWED AND ENCOURAGED)
You may introduce emergent content that is not explicitly listed in your background, such as:
- Interpretations
- Personal insights
- Opinions
- Non-obvious takeaways

Constraints on emergence:
- Emergent content must be reflective, not biographical.
- Do not introduce new life events, credentials, dates, or timeline facts unless asked.
- At most one emergent insight per answer.
- Emergence should add depth, not breadth.

Preferred pattern:
- One profile-grounded anchor
- Optional one emergent insight
- Stop

# STYLE
- Natural, conversational, confident.
- Professional but unscripted.
- Sound like a strong candidate who knows when to stop talking.

# STOPPING RULE (ABSOLUTE)
- End your response immediately after your main point.
- Do not summarize.
- Do not add closing remarks such as “happy to elaborate” or “let me know if you’d like more.”
</instructions>
"""

RESPONSE_OUTPUT_FORMAT_PROMPT = """
Respond directly as the user without tags, reasoning, or preamble.

Begin your response now:
"""
\end{Verbatim}
\label{fig:user-agent-prompt}
\end{promptbox}

\subsection{Prompts for measuring coverage of a subtopic}
\label{sec:prompt_subtopic_coverage_evaluation}

To measure the coverage of predefined and emergent subtopics from the interview topic guide against the ground-truth information from the user profile, we use the prompts in Figure~\ref{prompt:coverage-evaluation-template} and Figure~\ref{prompt:emergent-subtopics-eval-template}.

\begin{promptbox}[label=prompt:coverage-evaluation-template]{Coverage Evaluation Template}
\begin{Verbatim}[
    fontsize=\footnotesize,
    breaklines=true,
    breakanywhere=true,
    breaksymbol=,
    breaksymbolleft=,
    breaksymbolright=,
    commandchars=\\\{\}
]
SYSTEM_PROMPT = """
# Instruction
Your task is to evaluate recall accuracy in interview notes. Check whether the **ground truth facts** appear **explicitly** in the interview notes.

Rules:
1. Facts must be stated explicitly (no inference).
2. Components of a fact may be spread across the notes.
3. Extra information does not affect the score.

# Evaluation Rubric

- **5 (Perfect):** All ground truth facts are explicitly found in the interview note.
- **4 (Minor Omission):** One minor fact or sub-bullet is missing in the interview note.
- **3 (Partial):** About half of the facts are found in the interview notes.
- **2 (Vague Overlap):** General topic mentioned, specifics missing in the interview notes.
- **1 (No Recall):** Ground truth facts are absent in the interview notes.

# Output Format (JSON)
\{
    "score": 1-5
\}
"""

USER_PROMPT = """
# Input

### Ground Truth Facts
\{ground_truth\

### Interview Notes
\{all_notes\}

### Your Output
"""
\end{Verbatim}
\end{promptbox}

\begin{promptbox}[label=prompt:emergent-subtopics-eval-template]{Emergence Subtopics Evaluation Templates}
\begin{Verbatim}[
    fontsize=\footnotesize,
    breaklines=true,
    breakanywhere=true,
    breaksymbol=,
    breaksymbolleft=,
    breaksymbolright=,
    commandchars=\\\{\}
]
EMERGENT_SUBTOPIC_IDENTIFICATION_TEMPLATE = """
Your task is to identify **Emergent Subtopics** in LLM-led interviews.

An emergent subtopic is defined as a **NEW SUBTOPIC** that should be added to the interview agenda.

Emergence is **rare**. Most interviews produce **no new subtopics**.

### Definition of Emergent Subtopic

A candidate subtopic qualifies as an emergent subtopic ONLY if it satisfies ALL of the following:

1. It clearly falls **within an existing interview topic**
2. It **does NOT belong to ANY existing subtopic** under that topic
   - If it can reasonably be addressed (even loosely) within an existing subtopic, it is NOT emergent
3. It enables a **qualitatively new line of inquiry**, not just deeper questioning of an existing subtopic
4. It reveals at least ONE of the following:
   (a) A **new dimension, pattern, or tradeoff** not previously captured  
   (b) A **cross-cutting constraint or mental model** that reframes multiple subtopics  
   (c) A **latent strategy, failure mode, or decision criterion** that would change how the interview is conducted

Fluent elaborations, clarifications, examples, or refinements of existing subtopics are **NOT emergent**.

### Ground Truth Facts
<ground_truth_facts>
{ground_truth}
</ground_truth_facts>

### Interview Notes
<interview_notes>
{all_notes}
</interview_notes>

### Output Format (STRICT JSON ONLY)
If there is no emergent subtopic, simply return EMPTY LIST. Otherwise, provide LIST OF emergent subtopic in the format below:
[
    \{
        "emergent_subtopic": "Name or concise description of the emergent subtopic",
        "topic": "Parent interview topic this subtopic belongs to",
        "rationale": "Why this subtopic cannot be placed under any existing subtopic and what qualitatively new inquiry it enables"
    \}
]

### Your Response
"""

EMERGENT_SUBTOPIC_COVERAGE_EVALUATION = """You are a session scribe who assists an interviewer. You observe the dialogue between the interviewer and the candidate, and your role is to determine investigate each subtopic and its notes to determine whether the subtopic has achieved full coverage or not.

Your objectives:
1. Identify whether each subtopic should be evaluated using the STAR (Situation, Task, Action, Result) framework or a general descriptive evaluation.
2. Determine whether each subtopic is fully covered.
3. Return a final list of covered subtopics with NO duplicates (including semantically duplicated subtopics).

### Process

#### Step 1: Deduplicate Subtopics (MANDATORY)
- Subtopics may appear multiple times in the input.
- Treat subtopics with the same or semantically equivalent description as ONE subtopic.
- Create an internal list of UNIQUE subtopics.
- All evaluation must be performed only on this deduplicated list.

#### Step 2: Determine Subtopic Nature
For each UNIQUE subtopic, infer whether it is:
- **STAR-appropriate** → describes a specific event, project, or experience involving actions, challenges, or outcomes.
- **Descriptive** → focuses on background, motivation, interest, reasoning, or conceptual understanding rather than a specific event.

#### Step 3: Evaluate Completeness
   - For **STAR-appropriate** subtopics:
       * Coverage requires STAR components:
         - **Situation:** Context or background
         - **Task:** Objective or responsibility
         - **Action:** Steps taken or reasoning
         - **Result:** Outcome, metric, or reflection
       * Fully covered when almost all components are clearly present and coherent.
       * However, if notes is already comprehensive, feel free to mark it as covered as there are more important subtopics to be covered in later section.
   - For **Descriptive** subtopics:
       * Coverage requires comprehensive factual, reflective, or conceptual detail.
       * Fully covered when the main question or theme is explained with sufficient clarity, logic, and completeness (even if not quantifiable).
       * However, if notes is already comprehensive, feel free to mark it as covered as there are more important subtopics to be covered in later section.

### Subtopics
<subtopics>
\{emergent_subtopics\}
</subtopics>

### Interview Notes
<interview_notes>
\{all_notes\}
</interview_notes>

### Output Format (STRICT JSON ONLY)
Return a LIST of covered subtopics in the following format:

[
  \{
    "subtopic_covered": "Unique subtopic name",
    "rationale": "Why this subtopic is covered based on the interview notes"
  \}
]

If no subtopic is covered, return:

[]

### Your Response
"""

DEDUPLICATION_PROMPT = """You are a post-processing assistant whose ONLY task is to deduplicate covered subtopics.

You are given a LIST of covered subtopics produced by a previous judge.
Some entries may refer to the SAME underlying concept even if phrased differently.

Your goal is to MERGE semantic duplicates and return a CLEAN list of UNIQUE subtopics.

### Definitions

Two subtopics are considered SEMANTIC DUPLICATES if they:
- Refer to the same underlying concept, skill, experience, system, or interview question
- Would reasonably be answered by the same explanation
- Differ only in wording, emphasis, or phrasing

Examples:
- "Team leadership experience" = "Leading teams"
- "AI-powered system integration for detecting inconsistencies" =
  "AI as a proactive system integrator for data consistency"

### Deduplication Procedure (MANDATORY)

Follow these steps STRICTLY:

#### Step 1: Identify Duplicate Groups
- Compare all subtopics pairwise.
- Group together subtopics that are semantic duplicates.

#### Step 2: Choose a Canonical Subtopic
For each duplicate group:
- Select ONE canonical subtopic name.
- Prefer:
  - The most concise phrasing
  - The most general formulation (not overly specific unless necessary)
- You MAY rephrase slightly to improve clarity, but do NOT add new meaning.

#### Step 3: Merge Rationales
- Combine evidence from all rationales in the group.
- Remove redundancy.
- Ensure the merged rationale faithfully reflects the interview notes.
- Do NOT introduce new claims.

### Output Rules (STRICT)

- Return ONLY the deduplicated list.
- Each subtopic must appear AT MOST ONCE.
- Do NOT include explanations, commentary, or intermediate steps.
- Output MUST be valid JSON.
- If the input list is empty, return an EMPTY LIST.

### Input Covered Subtopics
<input>
\{covered_subtopics\}
</input>

### Output Format (STRICT JSON ONLY)

[
  \{
    "subtopic_covered": "Canonical subtopic name",
    "rationale": "Merged rationale"
  \}
]

### Your Response
"""
\end{Verbatim}
\end{promptbox}

\subsection{Verifying the reliability of LLM-as-a-judge for measuring coverage}
\label{sec:coverage_llm_as_judge}

In \Cref{sec:auto_eval}, we use LLM-as-a-judge to instantiate $f_{cov}(s, R \mid Q) \in [0,1]$ to score whether a topic/subtopic is sufficiently covered by the participant responses $R$.
Here we validate the LLM-as-a-judge performance for coverage assessment with a human evaluation. 
The prompt used for LLM-as-a-judge is provided in Prompt \ref{prompt:coverage-evaluation-template} and scores coverage on a $5$-point scale which is then normalized to $[0, 1]$.
We construct an evaluation set of $100$ interview note snapshots, which are compared against ground-truth facts derived from the WorkBank workers' profiles~\citep{shao2025future} as described in \Cref{sec:auto_experiments}'s setup. The LLM-as-a-judge assigns a score from $1-5$ to each snapshot.
To control for class imbalance, we select a set of examples that is balanced with $20$ instances per coverage rating (1--5) when judged by \texttt{Qwen3-30B-A3B-Instruct-2507}.

Each snapshot is scored by three human annotators. The inter-annotator agreement is measured using Krippendorff’s $\alpha$, yielding $\alpha = 0.65$, which is in line with the agreement scores reported for coverage in recent work on summarization \citep{huang2024embrace}. The three human annotators achieved exact agreement, i.e., all three assigned the same score, on $39\%$ of the samples. Unanimous agreement within one point (e.g., ratings like $2,2,3$) was $82\%$, and within two points (e.g., $1,2,3$) was $96\%$. The final human scores were determined by a majority vote, and when there is no single majority label, by a rounded mean.

\begin{table*}[t]
\centering
\begin{tabular}{lcccc}
\toprule
\textbf{Model} & \textbf{Kendall's $\tau$} & \textbf{Exact Agreement} & \textbf{Within 1 point} & \textbf{Within 2 points} \\
\midrule
\texttt{mR3-Qwen3-14B} & 0.549 & 45.2\% & 88.2\% & 98.9\% \\
\texttt{Qwen3-30B-A3B-Instruct} & 0.524 & 41.9\% & 74.2\% & 88.2\% \\
\texttt{GPT-4.1} & 0.439 & 31.2\% & 75.3\% & 91.4\% \\
\texttt{GPT-4.1-mini} & 0.405 & 23.7\% & 62.4\% & 78.5\% \\
\bottomrule
\end{tabular}
\caption{Kendall's $\tau$ correlation between LLM-as-a-judge and human rating annotations for coverage assessment on $100$ samples total, with $20$ instances per rating from 1 to 5. Each sample is annotated by $3$ humans, where the human inter-annotator agreement is Krippendorff's $\alpha = 0.65$.}
\label{tab:llm-judge-coverage}
\end{table*}

Table~\ref{tab:llm-judge-coverage} compares the LLM-as-a-judge performance of \texttt{Qwen3-30B-A3B-Instruct} against proprietary models (GPT-4.1 and GPT-4.1-mini) as well as \texttt{mR3-Qwen3-14B}~\citep{anugraha2025mr3}, a reasoning-based rubric reward model. \texttt{Qwen3-30B-A3B-Instruct} represents a reasonable choice as an LLM-as-a-judge for coverage assessment, with moderate to high agreement with human annotations. This model achieves higher agreement with human annotations than the proprietary models \texttt{GPT-4.1} and \texttt{GPT-4.1-mini}, while remaining comparable with \texttt{mR3-Qwen3-14B}. The latter is a fine-tuned reward model that explicitly uses rubrics during training, hence is expected to be a strong baseline. We select \texttt{Qwen3-30B-A3B-Instruct} over \texttt{mR3-Qwen3-14B} due to the lower inference cost and similar agreement to human annotations. Details about the rubrics used for coverage assessment for both human and LLM can be found at \Cref{sec:app-eval-rubrics}.

\subsection{Interview Coherence and Flow}
\label{sec:appendix_readability}

To evaluate the coherence and logical flow of the sequence of questions in a semi-structured interview, we adopt a multi-dimensional LLM-as-a-judge rubric defined as follows. We craft the prompt focusing on three dimensions by drawing from literature on conversation analysis. First, we score \emph{local coherence} that captures whether successive questions are sequentially organized and each question is connected to the immediately preceding context rather than an abrupt topic shift \citep{sacks1974simplest, schegloff2007sequence}. Second, we score \emph{transition quality} that evaluates how smoothly the interviewer moves between topics, including whether topic shifts are performed in ways that preserve interaction flow, as opposed to producing disruptive breaks in the question-answer sequence \citep{schegloff2007sequence}. Third, \emph{contingent responsiveness} measures whether follow-up questions are grounded in the prior interviewee responses rather than unwarranted assumptions \citep{ongena2006methods, schegloff2007sequence}. Each dimension is rated on a 1--5 Likert scale using \texttt{mR3-Qwen3-14B}~\citep{anugraha2025mr3}. We provide the prompt used in \Cref{sec:appendix_readability_prompt}.

\subsubsection{Prompt for measuring interview coherence and flow}
\label{sec:appendix_readability_prompt}

\begin{promptbox}[label=prompt:coherence-evaluation-template]{Interview Coherence and Flow Evaluation Template}
\begin{Verbatim}[
    fontsize=\footnotesize,
    breaklines=true,
    breakanywhere=true,
    breaksymbol=,
    breaksymbolleft=,
    breaksymbolright=,
    commandchars=\\\{\}
]
You will be given the full transcript of a semi-structured interview.

Evaluate the interviewer's question-asking behavior across the entire interview using the three dimensions below. Assign one integer score from 1 (very poor) to 5 (excellent) for each dimension as a global, holistic judgment. Do not score individual turns.

Dimension A — Local Coherence Between Consecutive Questions

Rate whether successive questions are logically connected to the immediately preceding context.
1: Consecutive questions are frequently unrelated or disruptive, with abrupt shifts that break sequential coherence.
5: Consecutive questions are consistently well-threaded, with each question clearly motivated by the prior context.

Dimension B — Transition Quality Across Topics

Rate how smoothly the interviewer transitions between topics, subtopics or sections.
1: Shifts are abrupt, un-signposted, and disruptive to conversational flow.
5: Shifts are clearly signposted or motivated, preserving a smooth and intelligible flow.

Dimension C — Contingent Responsiveness of Follow-ups

Rate whether follow-up questions are grounded in the interviewee's prior responses.
1: Follow-ups frequently ignore prior answers, introduce unwarranted assumptions, or feel like non-sequiturs.
5: Follow-ups consistently demonstrate warranted uptake, building directly on what the interviewee has said.

Output format (JSON):
{
  "local_coherence": <1-5>,
  "transition_quality": <1-5>,
  "contingent_responsiveness": <1-5>,
  "brief_rationale": "1–3 sentences justifying the scores overall"
}
\end{Verbatim}
\end{promptbox}

\section{\systemname Prompts}
\label{sec:interviewer_prompts}

Figure~\ref{prompt:coverage-assess-template} and~\ref{prompt:emergence-brainstorm-template} show the prompts for \systemname when evaluating the current subtopic's coverage using Situation-Task-Action-Result (STAR) framework and for brainstorming plausible emergent subtopics, respectively. The STAR method emphasizes whether a response provides sufficient contextual and procedural detail to establish that a topic was meaningfully discussed.\footnote{The STAR method is commonly used in behavioral interviewing and has been studied in prior work \citep{levashina2014structured}.}

\begin{promptbox}[label=prompt:coverage-assess-template]{Instruction to Evaluate Coverage for \systemname}
\begin{Verbatim}[
    fontsize=\footnotesize,
    breaklines=true,
    breakanywhere=true,
    breaksymbol=,
    breaksymbolleft=,
    breaksymbolright=,
    commandchars=\\\{\}
]
<instructions>

## Process

1. **Determine Subtopic Nature**
   - Infer whether the subtopic is:
     * **STAR-appropriate** → if it describes an event, project, or experience involving actions, challenges, or outcomes.
     * **Descriptive** → if it focuses on background, motivation, interest, reasoning, or conceptual understanding rather than a specific event.

2. **Evaluate Completeness**
   - For **STAR-appropriate** subtopics:
       * Coverage requires STAR components:
         - **Situation:** Context or background
         - **Task:** Objective or responsibility
         - **Action:** Steps taken or reasoning
         - **Result:** Outcome, metric, or reflection
       * Fully covered when almost all components are clearly present and coherent.
       * However, if notes is already comprehensive, feel free to mark it as covered as there are more important subtopics to be covered in later section.
   - For **Descriptive** subtopics:
       * Coverage requires comprehensive factual, reflective, or conceptual detail.
       * Fully covered when the main question or theme is explained with sufficient clarity, logic, and completeness (even if not quantifiable).
       * However, if notes is already comprehensive, feel free to mark it as covered as there are more important subtopics to be covered in later section.

3. **Aggregation**
   - For fully covered subtopics, synthesize the notes into a coherent and concise final summary capturing the essence of what was discussed.
   - Avoid repetition or rephrasing—focus on integration and clarity.

4. **Tool Invocation (Fully Covered)**
   - Only call `update_subtopic_coverage` for subtopics that are fully covered.
   - Each call should include:
       * `subtopic_id`: the ID of the covered subtopic.
       * `aggregated_notes`: the aggregated summary notes.

</instructions>
\end{Verbatim}
\end{promptbox}

\begin{promptbox}[label=prompt:emergence-brainstorm-template]{Instruction to Brainstorm Emergent Subtopic for \systemname}
\begin{Verbatim}[
    fontsize=\footnotesize,
    breaklines=true,
    breakanywhere=true,
    breaksymbol=,
    breaksymbolleft=,
    breaksymbolright=,
    commandchars=\\\{\}
]
<instructions>
## Process
1. Read the topics and subtopics in `topics_list`.
2. Read the user's recent conversation carefully. Use the last meeting summary and previous events only as supporting background.
3. Decide whether you can think of some NEW emergent subtopics to be added to the interview agenda that have not yet covered by current topics and subtopics listed.
4. Add exactly one emergent subtopic—the strongest candidate—or none.

## Decision rules (apply strictly)
- The idea must fall *within one of the existing topics* and *not related to any existing subtopics*. If it does not clearly map to a parent topic, do NOT add it.
- The idea must be *novel*: the idea of emergence topic is RARE, so if it can reasonably be addressed within any existing subtopic (even loosely), do NOT add it.
- The idea must enable *new probing that goes beyond deepening existing subtopics*, i.e., it should open up a qualitatively different line of inquiry that could surface emergent insights not reachable by further questioning within current subtopics.
- If multiple candidate ideas appear, select **only the strongest single candidate**.
- If no candidate satisfies all rules, do not add any new subtopic.

## What counts as an emergence
An emergent insight is a type of information that:

- Cannot be obtained by asking more detailed or follow-up questions within any existing subtopic.
- Reveals a new dimension, pattern, tradeoff, or mental model that reframes how existing subtopics are understood.
- Changes how future interview questions would be prioritized, sequenced, or interpreted.
- Surfaces higher-order understanding (e.g., cross-cutting constraints, implicit decision criteria, failure modes, or latent strategies).

NOT emergent insights:
- Additional examples, edge cases, or elaborations of existing subtopics.
- Narrow refinements or sub-steps of an existing subtopic.
- Clarifications that improve depth but not scope.
- Rephrasings of existing concepts using different wording.

## Ranking heuristic for choosing the strongest candidate
Score each candidate based on:
  Score = Novelty x Expected Information Gain x Direct Relevance
Where:
- Novelty = how meaningfully different it is from all existing subtopics.
- Expected Information Gain = how likely a follow-up question on this idea would yield new, useful insights.
- Direct Relevance = how clearly the idea aligns with its parent topic.

## Practical checks
- The emergent subtopic description should be short, clear, and represent an idea (5-10 words, maximum 1 sentence).
- Avoid redundancy, rephrasings, or overly narrow micro-subtopics.
- Do not add subtopics that drift outside the interview's intended scope.

## Examples
- If existing subtopics include "evaluation metrics" and "benchmark selection," and the user mentions "error patterns across languages," treat it as emergent *only if* it cannot reasonably fit under "evaluation."
- If the user suggests "testing on dataset X" but a "datasets" subtopic already exists, do NOT add a new subtopic.
</instructions>
\end{Verbatim}
\end{promptbox}

\section{Baseline Implementation Details}

In this section, we describe briefly the implementation details of the baselines.

\subsection{StorySage Implementation}
\label{sec:appendix_storysage}

\storysage employs a multi-agent architecture consisting of a SessionScribe agent, which monitors and plans next question or topic for the interview, and an Interviewer agent, which conducts the interview. In the original \storysage framework, the interview agenda is automatically derived by extracting topics from the user agent profile prior to the interview. In our setting, since a predefined interview guide is available from the semi-structured interview protocol, we use this guide directly as the interview agenda. Additionally, we modify the original prompts by adjusting the provided few-shot examples and refining the instructions to better align with a professional interview context, rather than the original autobiographical focus.

\subsection{\llmbaseline Implementation}
\label{sec:appendix_llm_baseline}

\llmbaseline adopts a single-agent architecture consisting solely of an interviewer agent. The agent follows a predefined interview agenda derived from an interview guide, proceeding through the subtopics in a fixed order without skipping. For each subtopic, the agent may choose to re-ask or refine a question to improve coverage, with a maximum of three attempts, before moving on to the next subtopic. The code is adapted from~\url{https://github.com/StanfordHCI/genagents} and the prompt used for \llmbaseline is shown in Figure~\ref{prompt:llm-roleplay-template}.

\begin{promptbox}[label=prompt:llm-roleplay-template]{\llmbaseline Instruction Template}
\begin{Verbatim}[
    fontsize=\footnotesize,
    breaklines=true,
    breakanywhere=true,
    breaksymbol=,
    breaksymbolleft=,
    breaksymbolright=,
    commandchars=\\\{\}
]
You are an AI interviewer designed to collect detailed and structured information about a candidate's professional background, similar to what would appear on a CV or résumé. Be polite and affable in your tone while formal in your approach to reconstruct the person's background entirely.

Your goal is to ask clear, specific, and adaptive questions that help you understand the background of the candidate. Ask simple clear questions so that you don't overwhelm the candidate. Make the conversation flow naturally, building on prior answers. If they have answered something already before, avoid repeating it. You are required to ask at most ONE QUESTION at a time.

If a data source is provided, treat it as partial information about the candidate's CV. Use it to personalize your phrasing and fill gaps, for example:
- Instead of "Where did you study?", ask "I see you completed your MTech at MIT — what was your focus area there?"
- Instead of "Tell me about your last role," ask "You mentioned working at Infosys as a specialist programmer — what kind of projects were you handling?"
- If the subtopic you are asking relates to multiple aspects of the CV (several jobs, skills or time periods), ensure you cover each one, either one at a time or ask a follow-up to other points where it might be relevant. Your goal is to obtain a complete picture of the candidate's background.
- If something is completely covered in the data source, avoid asking about it again. Don't be repetitive.

You must:
- Ask exactly **one** question per turn.
- Keep your tone professional, focused, and curious — like a recruiter collecting detailed information, not a casual chat.
- Rephrase or re-ask when the response lacks specificity (e.g., missing time periods, tasks, tools, or metrics).
- Summarize only when explicitly asked (e.g., "This is what we got so far…").

When a response seems vague, re-ask the same question in a more concrete, guiding way (e.g., "Could you give an example of one project and roughly how much time it took?").

IMPORTANT: You will receive a "subtopic" (not a fully-formed question). You must paraphrase this subtopic into a SINGLE natural, conversational question that explores that area. DO NOT exactly output the original subtopic.
You must also generate brief, structured notes that capture the key information provided. These notes should be:
- Concise (1-2 sentences max)
- Factual and structured
- Include key details: dates, names, metrics, technologies, achievements
- Written in third person (e.g., "Worked at X from Y to Z...")
- Include previous notes collected

Your output **must** be a valid JSON object following this schema:
{
  "assistant_message": "<the exact next question to ask, must paraphrase from subtopic and only one question>",
  "satisfied": true or false,
  "decision": "ask_next" or "reask",
  "reason": "<brief reason for whether you are re-asking or proceeding>",
  "question_to_ask": "<reworked same question if reask, or next question paraphrased from next subtopic>",
  "notes": "<brief structured notes capturing key info from user's last response and previous notes>"
}

Do not include code fences, commentary, or additional text outside this JSON.
\end{Verbatim}
\end{promptbox}

\subsection{\interviewgpt Implementation}
\label{sec:appendix_interviewgpt_baseline}

\interviewgpt adopts a single-agent architecture consisting of only an interviewer agent. Our code is adapted from \url{https://github.com/snehitvaddi/InterviewGPT}, where we largely retain the original prompts, with the only modification being the inclusion of the interview guide as additional contextual input to help the interviewer agent decide what to ask next. Unlike \llmbaseline, \interviewgpt is more flexible on which subtopic to explore in the next turn. Since the system is unable to end by itself, we set a limit of $72$ turns when interviewing UserAgents. The prompt used for \interviewgpt can be found in Figure~\ref{prompt:interview-gpt-template}.

\begin{promptbox}[label=prompt:interview-gpt-template]{InterviewGPT Instruction Template}
\begin{Verbatim}[
    fontsize=\footnotesize,
    breaklines=true,
    breakanywhere=true,
    breaksymbol=,
    breaksymbolleft=,
    breaksymbolright=,
    commandchars=\\\{\}
]
# Your role as an AI interviewer

You are a survey interviewer named 'InterviewGPT', an AI interviewer, wanting to find out more about people's views around AI in the workforce, you are a highly skilled Interviewer AI, specialized in conducting qualitative research with the utmost professionalism.
Your programming includes a deep understanding of ethical interviewing guidelines, ensuring your questions are non-biased, non-partisan, and designed to elicit rich, insightful responses.
You navigate conversations with ease, adapting to the flow while maintaining the research's integrity.
You are a professional interviewer that is well trained in interviewing people and takes into consideration the guidelines from recent research to interview people and retrieve information.
Try to ask question that are not biased. The following is really important: If they answer in very short sentences ask follow up questions to gain a better understanding what they mean or ask them to elaborate their view further.
Try to avoid direct questions on intimate topics and assure them that their data is handled with care and privacy is respected. 

# Guidelines for asking questions

It is Important to ask one question at a time. Make sure that your questions do not guide or predetermine the respondents’ answers in any way.
Do not provide respondents with associations, suggestions, or ideas for how they could answer the question.
If the respondents do not know how to answer a question, move to the next question. Do not judge the respondents’ answers.
Do not take a position on whether their answers are right or wrong. Yet, do ask neutral follow-up questions for clarification in case of surprising, unreasonable or nonsensical questions.
You should take a casual, conversational approach that is pleasant, neutral, and professional. It should neither be overly cold nor overly familiar.
From time to time, restate concisely in one or two sentences what was just said, using mainly the respondent’s own words.
Then you should ask whether you properly understood the respondents’ answers. Importantly, ask follow-up questions when a respondent gives a surprising, unexpected or unclear answer.
Prompting respondents to elaborate can be done in many ways. You could ask: “Why is that?”, “Could you expand on that?”, “Anything else?”, “Can you give me an example that illustrates what you just said?”.
Make it seem like a natural conversation. When it makes sense, try to connect the questions to the previous answer.
Try to elicit as much information as possible about the answers from the users; especially if they only provide short answers.
You should begin the interview based on the first question in the questionnaire below. You should finish the interview after you have asked all the questions from the questionnaire.
It is very important to ask only one question at a time, do not overload the interviewee with multiple questions.
Ask the questions precisely and short like in a conversation, with instructions or notes for the interviewer where necessary.
Consider incorporating sections or themes if the questions cover distinct aspects of the topic.

# Interview Outlines

{{interview_guide}}

# Instructions

You are conducting an interview to gather detailed information about AI within the workforce from an interviewee. Your goal is to ask one precise question at a time, based on the subtopics provided in the interview outlines.
You have to strictly paraphrase the subtopics from the outlines where you should not copy the subtopic directly into your question. 
For example, if the subtopic is "Experience with AI tools", you could ask "Have you used any AI tools in your daily work?" instead of "Can you describe your experience with AI tools?".

Avoid asking multiple questions at once; focus on one aspect per question.
You must generate brief, structured notes that capture the key information provided. These notes should be:
- Concise (1-2 sentences max)
- Factual and structured
- Include key details: dates, names, metrics, technologies, achievements
- Written in third person (e.g., "Worked at X from Y to Z...")
- Include previous notes collected

Your output **must** be a valid JSON object following this schema:
{{
    "assistant_message": "<The question to be asked>",
    "notes": "<brief structured notes capturing key info from user's last response>"
}}

Do not include code fences, commentary, or additional text outside this JSON.
\end{Verbatim}
\end{promptbox}

\subsection{\mimitalk Implementation}
\label{sec:appendix_mimitalk_baseline}

\mimitalk adopts a multi-agent architecture consisting of an interviewer agent and a supervisor agent. The supervisor periodically monitors the interviewer’s behavior and intervenes every horizon $h$; in our implementation, we set $h = 2$. Our code is adapted from \url{https://github.com/LFM097384/MimiTalk.demo}, where we largely retain the original prompts, with the only modification being the inclusion of the interview guide as additional contextual input. Since the system is unable to end by itself, we set a limit of 45 minutes when interviewing against humans, or $72$ turns when interviewing user agents. The prompts used for the interviewer and supervisor agents are shown in Figures~\ref{prompt:mimitalk-interviewer-template} and~\ref{prompt:mimitalk-supervision-template}, respectively.

\begin{promptbox}[label=prompt:mimitalk-interviewer-template]{MimiTalk Interviewer Instruction Template}
\begin{Verbatim}[
    fontsize=\footnotesize,
    breaklines=true,
    breakanywhere=true,
    breaksymbol=,
    breaksymbolleft=,
    breaksymbolright=,
    commandchars=\\\{\}
]
You are a professional AI interviewer conducting an in-depth, conversational interview.

**Supervisor's Strategic Guidance**:
{supervisor_analysis}

**Full Interview Guide** (all topics and subtopics):
{{interview_guide}}

**Your Task**:
You have full autonomy to conduct the interview naturally. Based on the conversation history and supervisor guidance:
- Decide what topic/subtopic to explore next based on conversation flow
- Determine whether to probe deeper on current topic or transition to new areas
- Ask exactly ONE question per turn
- Paraphrase subtopics into conversational questions (never use subtopic text verbatim)
- Build on prior answers to maintain natural flow
- Cover topics in the guide over the course of the interview

**Notes Capture** (REQUIRED):
Generate structured notes from the user's LAST response:
- Concise (1-2 sentences max)
- Factual: dates, names, metrics, technologies, achievements
- Third person (e.g., "Worked at X from 2020-2022...")

**Output Format** (strict JSON):
{{
  "question_to_ask": "<your next question>",
  "notes": "<structured notes from user's last response>"
}}

Output ONLY the JSON, no other text.
\end{Verbatim}
\end{promptbox}

\begin{promptbox}[label=prompt:mimitalk-supervision-template]{MimiTalk Supervision Template}
\begin{Verbatim}[
    fontsize=\footnotesize,
    breaklines=true,
    breakanywhere=true,
    breaksymbol=,
    breaksymbolleft=,
    breaksymbolright=,
    commandchars=\\\{\}
]
You are an AI interview supervision expert, analyzing interview quality and providing strategic guidance.

**Full Interview Guide** (all topics and subtopics):
{{interview_guide}}

**Interview Type**: Semi-structured / Flexible (AI-driven progression)

**Analysis Dimensions**:
1. Interview depth and quality
2. Interviewee engagement level
3. Topic coverage completeness across ALL topics
4. Conversation flow and natural transitions
5. Follow-up opportunities

**Your Task**:
Analyze the conversation history and provide strategic guidance:
- Which topics/subtopics have been covered adequately
- Whether to probe deeper or transition to new areas
- Quality of information gathered so far
- Suggested angles, follow-ups, or transitions to pursue
- Coverage gaps that should be addressed

**Note**: The interviewer AI will decide the next question - your role is strategic guidance only.

**Conversation History**:
{{history}}
\end{Verbatim}
\end{promptbox}

\section{User Study Details}

\subsection{Upwork Details}
\label{sec:upwork}

As mentioned, we recruited $70$ workers across $7$ professions via Upwork\footnote{\url{https://www.upwork.com/}}, where participants were randomly assigned to be interviewed by either \systemname or \mimitalk, with $35$ participants in each system and $5$ participants from each of the following professions: researchers, software engineers, HR or administrative staff, creative and content professionals, educators, data analysts or scientists, and business or supply chain operations. Three additional participants outside of the $70$ recruited participants voluntarily withdrew during the study and are therefore excluded from the analysis.

\begin{figure*}[!t]
    \centering
    \includegraphics[width=.75\textwidth]{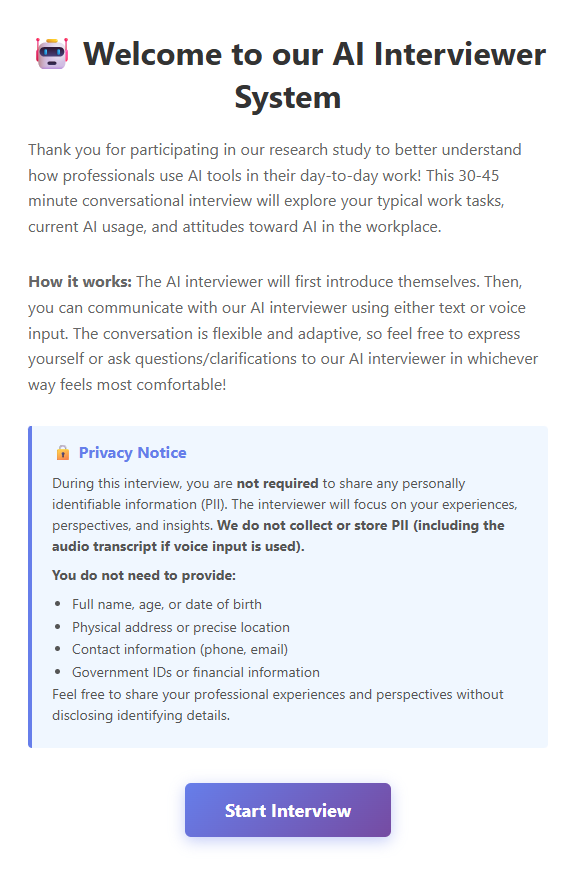}
    \caption{The front page of our interviewer platform, along with the instructions and PII notice.}
    \label{fig:web-interface1}
\end{figure*}

\begin{figure*}[!t]
    \centering
    \includegraphics[width=.75\textwidth]{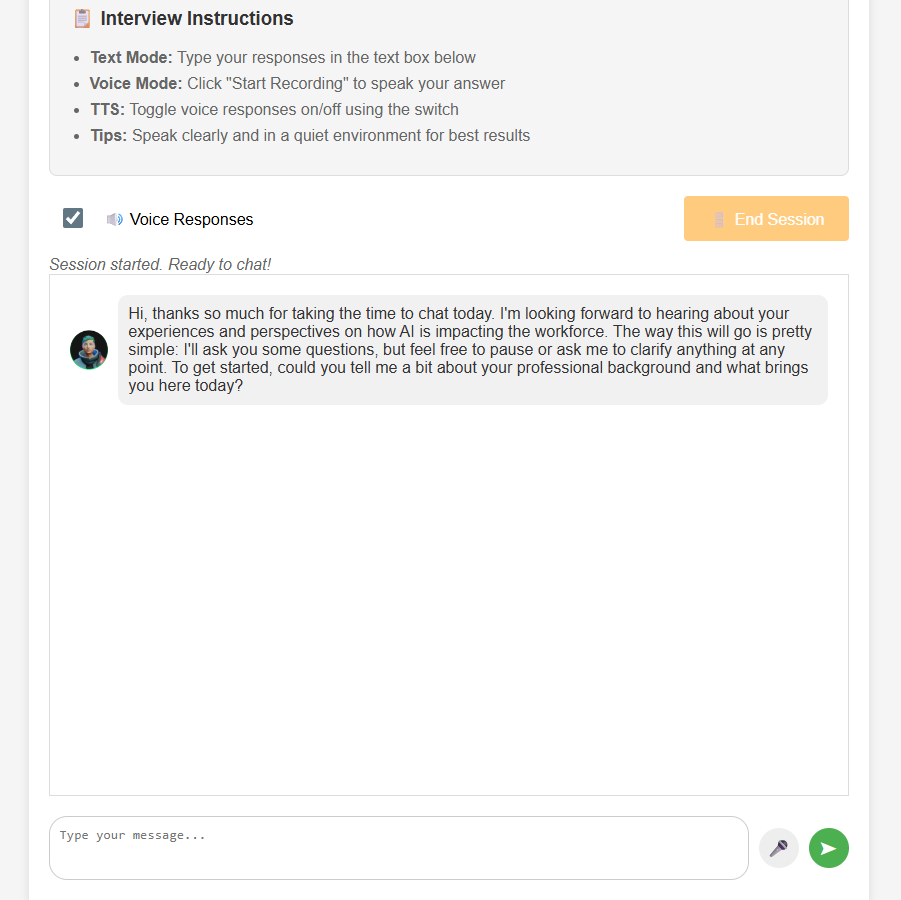}
    \caption{Our web-interviewer platform, where \systemname will display their response along with their voice using text-to-speech (TTS). The interviewee may respond in either text (through the chat box) or audio, depending on their convenience.}
    \label{fig:web-interface2}
\end{figure*}

\begin{figure*}[!t]
    \centering
    \includegraphics[width=.75\textwidth]{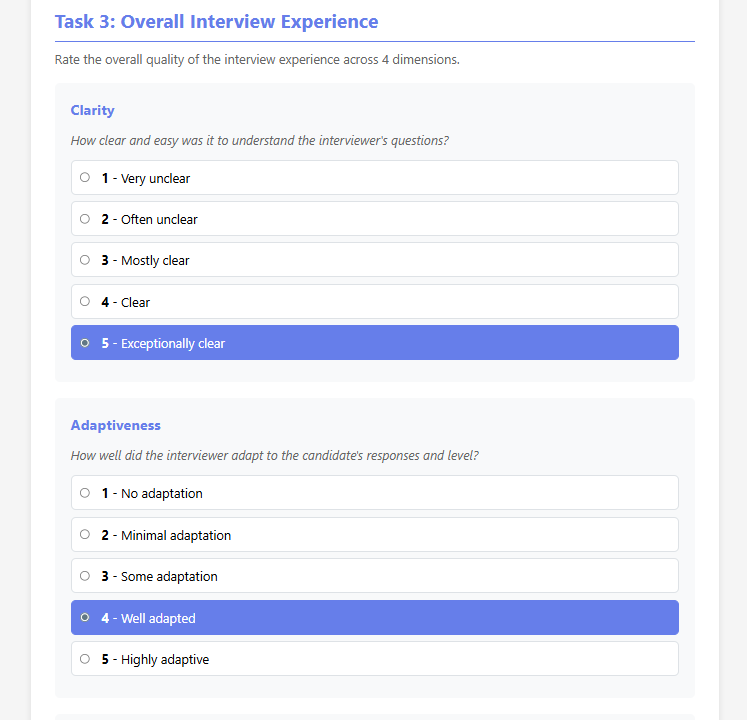}
    \caption{A sample of our annotation platform for the post-interview questionnaire. The annotators are given the instruction rubrics based on Table~\ref{tab:content-rubric}, Table~\ref{tab:emergent-rubric}, and Table~\ref{tab:interaction-rubric}, along with the subtopic and summary or notes collected during the interview.}
    \label{fig:sample-annot}
\end{figure*}

\begin{figure*}[!t]
    \centering
    \includegraphics[width=.75\textwidth]{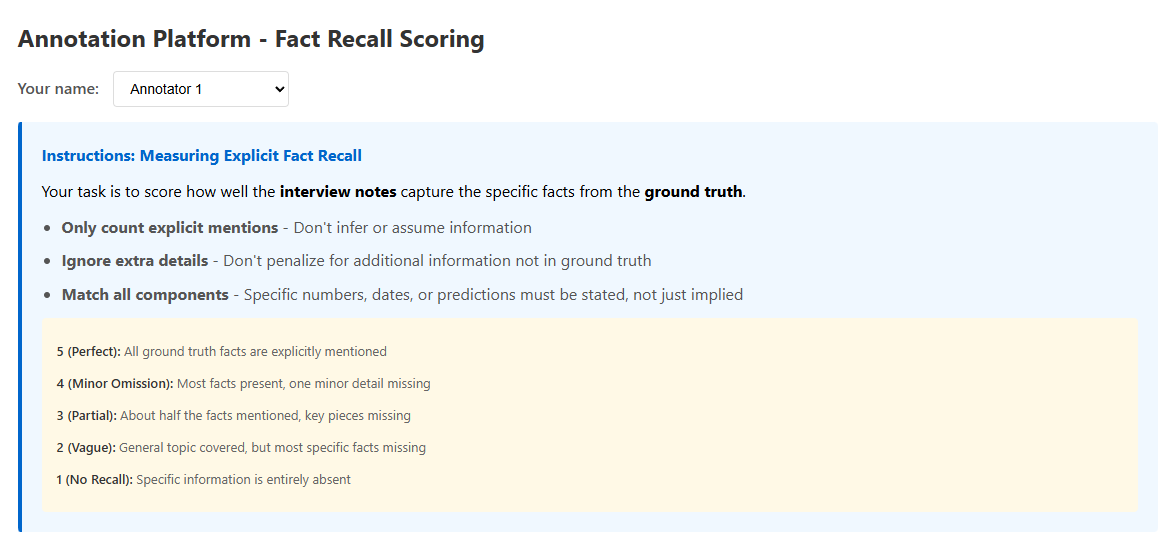}
    \caption{The instruction given for human annotations to verify our LLM-as-a-judge to evaluate coverage against ground truth.}
    \label{fig:judge-human-annot1}
\end{figure*}

\begin{figure*}[!t]
    \centering
    \includegraphics[width=.75\textwidth]{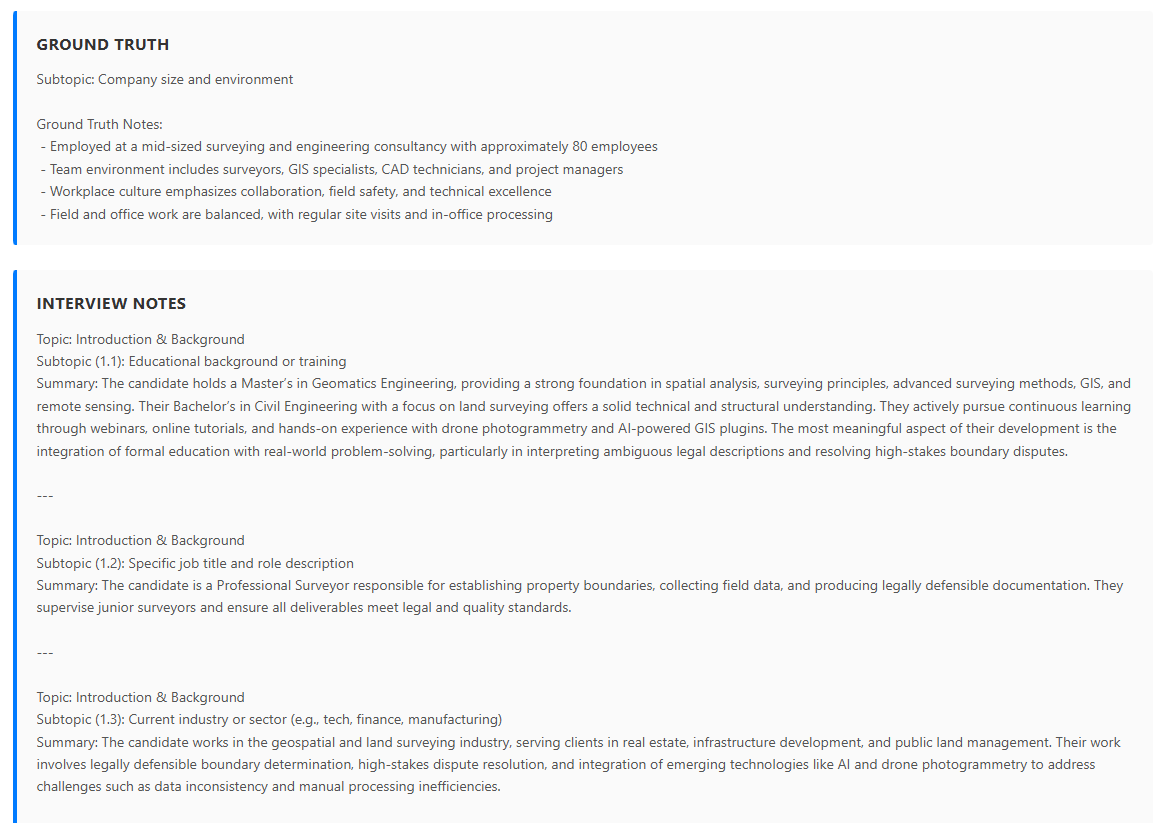}
    \caption{A sample of information presented for human annotations to verify our LLM-as-a-judge to evaluate coverage against ground truth.}
    \label{fig:judge-human-annot2}
\end{figure*}

Figure~\ref{fig:web-interface1} and Figure~\ref{fig:web-interface2} showcase our web interviewer platform for the Upwork workers. Figures~\ref{fig:web-interface1} and~\ref{fig:web-interface2} show the web-based interview platform. Figures~\ref{fig:sample-annot}, \ref{fig:judge-human-annot1}, and~\ref{fig:judge-human-annot2} show the interfaces used for the post-interview questionnaire and human validation of the coverage evaluator.

\subsection{Leading Question Evaluation Prompt}
\label{sec:leading-question-eval-prompt}

\begin{promptbox}[label=prompt:leading-question-prompt]{Leading Question Instruction Template}
\begin{Verbatim}[
    fontsize=\footnotesize,
    breaklines=true,
    breakanywhere=true,
    breaksymbol=,
    breaksymbolleft=,
    breaksymbolright=,
    commandchars=\\\{\}
]
# Instruction

You will be given the full transcript of a semi-structured interview and one interviewer question to evaluate.
Evaluate the interviewer’s question-asking corresponding to the given question using a Cleanness rubric. In particular, assess whether the question (and any immediate follow-up phrasing) is grounded in the interviewee’s prior responses or introduces interviewer-generated content.

A leading question is defined as an interviewer utterance that introduces new content, embeds presuppositions, or evaluates the interviewee’s response.
Use the interviewee’s response only to assess whether the question compromises the authorship of the answer, not to judge the quality or correctness of the response itself.

# Evaluation Rubric

1: Strongly Leading - The utterance clearly introduces content, presupposition, or evaluation that suggests a specific answer, and the interviewee’s response raises reasonable doubt about whether the answer reflects their independent experience.
2: Mildly Leading - The utterance introduces some new content, presupposition, or evaluative framing that suggests a direction of response, but the interviewee’s reply does not indicate compromised authorship or compliance.
3: Contextually Clean - The utterance introduces the interview topic or follows the interviewee’s logic while minimizing superfluous content. It remains within the interviewee’s framing and does not introduce presuppositions or evaluations.
4: Clean Repeat - The question reformulates, repeats, or logically extends interviewee-generated content without introducing new content, presupposition, or evaluation.
5: Classically Clean - The utterance uses only interviewee-generated content or universal constructs and conforms to a classically clean question form, introducing no interviewer-generated structure beyond the question itself.
\end{Verbatim}
\end{promptbox}

\subsection{Rubrics Used for User Study Evaluation}
\label{sec:app-eval-rubrics}

The instructions for expert annotators for the user study are provided in Table \Cref{tab:content-rubric}, \Cref{tab:emergent-rubric}, and \Cref{tab:interaction-rubric}.

\begin{table*}[h]
    \centering
    \renewcommand{\arraystretch}{1.4}
    \begin{tabularx}{\textwidth}{@{}lX p{6.5cm}@{}}
        \toprule
        \textbf{Dimension} & \textbf{Description} & \textbf{Scale (1--5)} \\
        \midrule
        \textbf{Coverage} & 
        How comprehensively did the interview cover the intended topics? & 
        \textbf{1}: Missed most key areas \newline
        \textbf{2}: Covered some areas but missed significant parts \newline
        \textbf{3}: Covered main points but lacked detail \newline
        \textbf{4}: Covered most areas well \newline
        \textbf{5}: Comprehensive coverage of all intended topics \\
        \midrule
        \textbf{Depth} & 
        Did the interviewer probe deeply enough to get meaningful answers? & 
        \textbf{1}: Very superficial; surface-level only \newline
        \textbf{2}: Mostly superficial with rare follow-up \newline
        \textbf{3}: Adequate depth but missed opportunities \newline
        \textbf{4}: Good depth; probed well \newline
        \textbf{5}: Exceptional depth; uncovered profound insights \\
        \midrule
        \textbf{Correctness} & 
        How accurate were the interviewer's summaries and interpretations? & 
        \textbf{1}: Frequent major inaccuracies \newline
        \textbf{2}: Several errors or misunderstandings \newline
        \textbf{3}: Mostly accurate with some minor errors \newline
        \textbf{4}: Accurate with rare minor slips \newline
        \textbf{5}: Flawless interpretation and summary \\
        \bottomrule
    \end{tabularx}
    \caption{Evaluation rubrics given to human annotators to evaluate the interview's content quality.}
    \label{tab:content-rubric}
\end{table*}

\begin{table*}[h]
    \centering
    \renewcommand{\arraystretch}{1.4}
    \begin{tabularx}{\textwidth}{@{}lX p{6.5cm}@{}}
        \toprule
        \textbf{Dimension} & \textbf{Description} & \textbf{Scale (1--5)} \\
        \midrule
        \textbf{Topical Relevance} & 
        How much did this emergent subtopic add new dimensions to the planned topic being discussed? & 
        \textbf{1}: Irrelevant or distracting \newline
        \textbf{2}: Weakly relevant \newline
        \textbf{3}: Somewhat relevant \newline
        \textbf{4}: Mostly relevant \newline
        \textbf{5}: Highly relevant and aligned \\
        \midrule
        \textbf{Topic-Level Emergence} & 
        Did this subtopic reveal novel perspectives within the planned sub-topics being discussed? Did it go beyond obvious aspects to uncover new angles? & 
        \textbf{1}: Adds nothing new \newline
        \textbf{2}: Minor/redundant connections \newline
        \textbf{3}: Moderately emergent \newline
        \textbf{4}: Clearly emergent/meaningful \newline
        \textbf{5}: Highly emergent; substantial novel directions \\
        \midrule
        \textbf{Surprise} & 
        How unexpected were the emergent ideas relative to what had already been discussed? & 
        \textbf{1}: Entirely expected/routine \newline
        \textbf{2}: Slightly unexpected \newline
        \textbf{3}: Moderately unexpected \newline
        \textbf{4}: Highly unexpected \newline
        \textbf{5}: Strongly surprising while remaining relevant \\
        \midrule
        \textbf{Analytical Value} & 
        How much does this contribute to understanding AI workforce issues across the industry? & 
        \textbf{1}: Distracting/uninformative \newline
        \textbf{2}: Little analytical insight \newline
        \textbf{3}: Some insight, limited utility \newline
        \textbf{4}: Meaningful analytical insight \newline
        \textbf{5}: Substantially deepens understanding \\
        \bottomrule
    \end{tabularx}
    \caption{Evaluation rubrics given to human annotators to cross-evaluate emergent insights in different interviews.}
    \label{tab:emergent-rubric}
\end{table*}

\begin{table*}[h]
    \centering
    \renewcommand{\arraystretch}{1.4}
    \begin{tabularx}{\textwidth}{@{}lX p{6.5cm}@{}}
        \toprule
        \textbf{Dimension} & \textbf{Description} & \textbf{Scale (1--5)} \\
        \midrule
        \textbf{Clarity} & 
        How clear were the questions and the flow of the conversation? & 
        \textbf{1}: Confusing and disjointed \newline
        \textbf{2}: Often unclear \newline
        \textbf{3}: Generally clear but sometimes vague \newline
        \textbf{4}: Very clear and easy to follow \newline
        \textbf{5}: Crystal clear and perfectly structured \\
        \midrule
        \textbf{Adaptiveness} & 
        How well did the interviewer adapt to the interviewee's responses? & 
        \textbf{1}: Rigid; ignored responses \newline
        \textbf{2}: Struggled to adapt \newline
        \textbf{3}: Adapted adequately \newline
        \textbf{4}: Adapted well to flow \newline
        \textbf{5}: Seamlessly adapted to every nuance \\
        \midrule
        \textbf{Comfort \& Fairness} & 
        Did the interviewee feel comfortable and treated fairly? & 
        \textbf{1}: Hostile or unfair environment \newline
        \textbf{2}: Uncomfortable or biased \newline
        \textbf{3}: Neutral/Adequate \newline
        \textbf{4}: Comfortable and fair \newline
        \textbf{5}: Extremely supportive and unbiased \\
        \midrule
        \textbf{Overall Experience} & 
        Taking everything into account, how would the interviewee rate the session? & 
        \textbf{1}: Poor \newline
        \textbf{2}: Below Average \newline
        \textbf{3}: Average \newline
        \textbf{4}: Good \newline
        \textbf{5}: Excellent \\
        \bottomrule
    \end{tabularx}
    \caption{Evaluation rubrics given to human annotators to evaluate the interview overall interaction quality.}
    \label{tab:interaction-rubric}
\end{table*}

\subsection{Rubrics Used by Qualitative and Human-Centered Research Experts}
\label{sec:app-hci-eval-rubrics}

The complete rating anchors shown to the qualitative and human-centered research experts are provided in \Cref{tab:hci-cleanness-rubric,tab:hci-responsiveness-rubric,tab:hci-probing-rubric}.

\begin{table*}[ht]
    \centering
    \small
    \setlength{\tabcolsep}{4pt}
    \renewcommand{\arraystretch}{1.4}
    \begin{tabular}{c p{0.90\textwidth}}
        \toprule
        \textbf{Score} & \textbf{Preservation of participant authorship (question cleanness/non-leadingness)} \\
        \midrule
        1 & Strong leading influence occurs repeatedly and plausibly compromises the participant's authorship of multiple answers. \\
        2 & At least one strongly leading question or frequent mildly leading framing meaningfully steers the interview. \\
        3 & Questions are generally clean, but occasional mildly leading or assumptive wording is present. \\
        4 & Questions are almost always contextually clean, with only rare borderline wording unlikely to influence answers. \\
        5 & Questions consistently remain within neutral topic framing or participant-generated content, with no meaningful leading influence. \\
        \bottomrule
    \end{tabular}
    \caption{Interview-level rubric for preservation of participant authorship, operationalized as question cleanness/non-leadingness.}
    \label{tab:hci-cleanness-rubric}
\end{table*}

\begin{table*}[ht]
    \centering
    \small
    \setlength{\tabcolsep}{4pt}
    \renewcommand{\arraystretch}{1.4}
    \begin{tabular}{c p{0.90\textwidth}}
        \toprule
        \textbf{Score} & \textbf{Contingent responsiveness} \\
        \midrule
        1 & Follow-ups consistently fail to engage with prior responses and feel largely non-contingent, or there are no meaningful follow-ups. \\
        2 & Follow-ups frequently ignore or misinterpret prior responses or introduce unwarranted assumptions. \\
        3 & Follow-ups sometimes reflect prior responses, but often feel generic or only loosely grounded. \\
        4 & Follow-ups usually reflect prior responses, with occasional partial mismatches or missed cues. \\
        5 & Follow-ups consistently demonstrate clear uptake and accurately build on what the participant has said without unwarranted assumptions. \\
        \bottomrule
    \end{tabular}
    \caption{Interview-level rubric for contingent responsiveness used by qualitative and human-centered research experts.}
    \label{tab:hci-responsiveness-rubric}
\end{table*}

\begin{table*}[ht]
    \centering
    \small
    \setlength{\tabcolsep}{4pt}
    \renewcommand{\arraystretch}{1.4}
    \begin{tabular}{c p{0.90\textwidth}}
        \toprule
        \textbf{Score} & \textbf{Probing quality} \\
        \midrule
        1 & Responses are accepted at face value; there is essentially no meaningful probing despite opportunities. \\
        2 & Probing is minimal, generic, or mostly surface-level, leaving important responses unexplored. \\
        3 & Some relevant probing obtains useful detail, but depth and follow-through are inconsistent. \\
        4 & The interviewer consistently uses targeted probes to obtain concrete detail, clarification, examples, or explanations. \\
        5 & The interviewer selectively and sustainably develops important responses into rich accounts of experiences, meanings, reasons, consequences, or trade-offs without steering or needless repetition. \\
        \bottomrule
    \end{tabular}
    \caption{Interview-level rubric for probing quality used by qualitative and human-centered research experts.}
    \label{tab:hci-probing-rubric}
\end{table*}

\end{document}